\documentclass[preprint,5p,sort&compress]{elsarticle}
\biboptions{sort&compress}

\usepackage{graphicx,wrapfig,lipsum}
\usepackage{mathtools}
\usepackage{mathrsfs} 
\usepackage{amsmath,amssymb,bm,accents}
\usepackage{amsthm}
\usepackage[version=4]{mhchem}
\usepackage{chemformula}
\usepackage{siunitx}
\usepackage{booktabs} 

\numberwithin{equation}{section}

\makeatletter
\def\ps@pprintTitle{%
 \let\@oddhead\@empty
 \let\@evenhead\@empty
 \def\@oddfoot{}%
 \let\@evenfoot\@oddfoot}
\makeatother

\newcommand\restr[2]{{
  \left.\kern-\nulldelimiterspace 
  #1 
  \vphantom{\big|} 
  \right|_{#2} 
  }}
\def\restrict#1{\raise-.5ex\hbox{\ensuremath|}_{#1}}

    \DeclareMathSymbol{\widetildesym}{\mathord}{largesymbols}{"65}

\DeclareMathOperator{\diag}{diag}
 
\begin{document}
\title{Structural electroneutrality in Onsager--Stefan--Maxwell transport with charged species}
\date{\today}
\author[maths,FI]{Alexander Van-Brunt}
\author[maths]{Patrick E. Farrell}
\author[engsci,FI]{Charles W. Monroe}
\ead{charles.monroe@eng.ox.ac.uk}
\address[maths]{Mathematical Institute, University of Oxford, Woodstock Road, Oxford, OX2 6GG, UK}
\address[engsci]{Department of Engineering Science, University of Oxford, Oxford, OX1 3PJ, United Kingdom}
\address[FI]{The Faraday Institution, Harwell Campus, Didcot, OX11 0RA, United Kingdom}

\begin{abstract}
We present a method to embed local electroneutrality within Onsager--Stefan--Maxwell electrolytic-transport models, circumventing their formulation as differential systems with an algebraic constraint. Flux-explicit transport laws are formulated for general multicomponent electrolytes, in which the conductivity, component diffusivities, and transference numbers relate to Stefan--Maxwell coefficients through invertible matrix calculations. A construction we call a `salt--charge basis' implements Guggenheim's transformation of species electrochemical potentials into combinations describing a minimal set of neutral components, leaving a unique combination associated with electricity. Defining conjugate component concentrations and fluxes that preserve the structures of the Gibbs function and energy dissipation retains symmetric Onsager reciprocal relations. The framework reproduces Newman's constitutive laws for binary electrolytes and the Pollard--Newman laws for molten salts; we also propose laws for salt solutions in two-solvent blends, such as lithium-ion-battery electrolytes. Finally, we simulate a potentiostatic Hull cell containing a non-ideal binary electrolyte with concentration-dependent properties.  
 \end{abstract}

\maketitle

\section{Introduction}
The growing impetus for improved electrochemical energy storage makes it vital to understand ion transport. Concentrated solution theory \cite{newman2004electrochemical, Newman1965} is widely used to model mass and heat transport in electrolytic materials on the continuum scale. One of the most popular implementations is Doyle, Fuller and Newman's landmark physics-based model of the dual-insertion lithium-ion battery, which incorporates a concentrated-solution model of a binary electrolyte, comprising a simple salt dissolved in an uncharged solvent \cite{Doyle_1993, Doyle01061996}.

Concentrated solution theory is grounded in both classical thermodynamics and linear irreversible thermodynamics, with a model structure that sharply distinguishes between equilibrium properties and transport properties. It derives from the Onsager--Stefan--Maxwell transport laws \cite{hirschfelder1954molecular, newman2004electrochemical}, written for $n$ species in an isotropic, isothermal, isobaric common phase as 
\begin{equation} \label{OSMequation}
- \nabla \mu_{i} = \sum_{j=1}^{n} M_{ij} \vec{N}_{j},
\end{equation}
where $\mu_i$ signifies the electrochemical potential of species $i \in \left\{ 1, ..., n \right\}$ and $\vec{N}_{i}$, its total molar flux. The transport coefficients make up an $n \times n$ matrix $\textbf{M}$ whose entries are
\begin{equation} \label{transport matrix}
M_{ij} = \begin{cases} -\dfrac{RT }{c_{\text{T}} \mathscr{D}_{ij} }  \:\:\: \text{if} \:\: i \neq j \vspace{6pt} \\
\dfrac{RT}{c_{\text{T}} } \displaystyle{ \sum_{k \neq i}^{n}} \dfrac{c_{k}}{\mathscr{D}_{ik} c_{j}} \:\:\: \text{if} \:\: i = j, \end{cases}
\end{equation}
in which $R$ represents the gas constant, $T$, the absolute temperature, and $c_{\text{T}}$, the total species molarity; $c_{i}$ is the molarity of species $i$, and $\mathscr{D}_{ij}$, the Stefan--Maxwell diffusivity of species $i$ through species $j$. It has been shown for solutions of uncharged constituents that Onsager's reciprocal relations and the second law of thermodynamics require $\textbf{M}$ to be symmetric positive semidefinite \cite{Monroe2015}. Also, $\textbf{M}$ affords a single null eigenvector $\textbf{c} =  [c_{1},c_{2},..., c_{n}]^\top$, a further structure which ensures that convection of material at the bulk velocity does not dissipate energy \cite{Onsager1945}. (In this work bold lower-case letters always indicate column matrices.) For fluids wholly made up of uncharged species, the Onsager--Stefan--Maxwell equations, when combined with an equation that establishes a convective velocity, have been shown to constitute a closed model, neither underdetermined nor overdetermined numerically \cite{VanbruntIMA}. 

An important practicality emerges when considering transport in liquids such as electrolytic solutions, which conduct electricity via the motion of ions. When species that carry charge are mobile, their thermal motion screens local charge imbalances on a length scale comparable to the Debye length (typically of the order of 1 nm) \cite{DH1}. Thus, if volume elements in a simulation have characteristic size much larger than the Debye length, the local excess charge density $\rho_\textrm{e}$ within them can be assumed to vanish locally, that is,
\begin{equation} \label{electroneutrality.appx}
\rho_{\textrm{e}} = F \sum_{j=1}^{n} z_{j} c_{j} = 0,
\end{equation}
where $F$ is Faraday's constant and $z_i$ is the equivalent charge of species $i$. Local electroneutrality is foundational to concentrated solution theory and has been assumed nearly ubiquitously in its applications. 

Adopting the local electroneutrality approximation simplifies transport models by allowing one charged-species concentration to be eliminated from the governing equations, both reducing the complexity of thermodynamic parametrization and streamlining calculations. Electroneutrality makes concentrated solution theory hard to reconcile with the Onsager--Stefan--Maxwell theory for nonelectrolytes, however, because the additional algebraic constraint imposed by equation \eqref{electroneutrality.appx} appears to induce an overdetermination within equations \eqref{OSMequation}. In particular, it is not clear whether the symmetry and rank of the transport matrix are preserved within electrolytic systems after electroneutrality restricts the available space of species compositions.
 
Newman and colleagues carefully circumvented the issues raised by electroneutrality when formulating models of binary electrolytes and salt melts \cite{Newman1965,NewmanChapman1973,Pollard_1979}. These analyses emphasized the importance of separating terms into neutral and non-neutral contributions: individual ion concentrations were lumped into salt concentrations; species fluxes were expressed in terms of salt chemical potentials and current density; and terms associated with charge density and charge convection were explicitly neglected.  

Possible applications of electroneutral Onsager--Stefan--Maxwell theory readily exceed the scope of the few cases that have been considered to date. For example, aqueous vanadium flow batteries---a transport system comprising at least eight distinct species---have recently been modelled \cite{Adamflowbattery2, Adamflowbattery3}, but it is not immediately clear how such a model could be extended to account for additional components in a consistent way. At the same time, the increased demand for battery simulations that account for electrolyte additives or contaminants, solution-phase reaction intermediates, or products of side reactions raises the need for a general, numerically robust model formulation. 
 
This paper generalizes Newman's particular implementations of electroneutral concentrated solution theory for binary solutions and salt melts. We implement Guggenheim's thermodynamic programme \cite{Guggenheim} by creating a structure-preserving, invertible linear transformation that maps the free-energy changes incurred by individual charged species into changes involving uncharged combinations of species, assembling the species electrochemical potentials into a basis set of component chemical potentials and a single quantity that can be interpreted as an electric potential. The same transformation regroups charged-species fluxes into component fluxes and current density. As well as reconciling the electroneutrality approximation with the fundamental equations of linear irreversible thermodynamics in a structure-preserving way, the theory produces a general set of transference numbers and an ionic conductivity that match definitions given for binary electrolytes by Newman, as well as leading to a naturally symmetric set of component diffusivities.

Our development enables the extension of Newman's flux-explicit transport laws for binary electrolytes to general multicomponent electrolytes containing any number of charged or uncharged species, while maintaining the crucial benefits of thermodynamic consistency, convection-invariance of diffusion, and symmetric reciprocal relations inherited from the Onsager--Stefan--Maxwell equations. A numerical simulation of a Hull cell that leverages a solver for Onsager--Stefan--Maxwell transport illustrates the practical utility of this structure-based approach to local electroneutrality within concentrated solution theory. 

\section{Construction of a salt--charge basis}
\label{sec:construction}

The key mathematical concepts employed below are linear transformations and inner products, typically represented by matrix operations.  These tools are used to develop a change of composition basis from one that lists charged-species molarities into one that lists neutral-salt molarities alongside charge density. Instead of beginning with species concentrations, however, the new composition basis is identified by following Guggenheim's process of grouping species electrochemical potentials into new quantities that represent electrically neutral combinations of species \cite{Guggenheim}. For any given electrolytic material, one can establish a minimal set of such combinations by positing a set of independent hypothetical chemical reactions that we call \emph{simple association equilibria}, which leave uncharged species alone or bring charged species together to form neutral salts. We refer to products of the simple association equilibria as \emph{components}, rather than species. 

Every electrochemical system is globally electroneutral at equilibrium and every electrochemical reaction necessarily balances both charge and atoms between its reactants and products. Moreover, any physically realizable electrolytic solution must contain at least two countercharged species. It follows that the $n$ species comprising any electrolyte combine to form at most $n-1$ distinct uncharged components. Thus, association equilibria can be used to combine ions into a set of uncharged products of association reactions, whose energetics in solution is described unambiguously in terms of classical chemical potentials. This idea dates back to the work of Guggenheim almost a century ago \cite{Guggenheim}; the process we outline in this paper is merely intended to systematize Guggenheim's framework, and consequently, to enable an analysis of electrolyte thermodynamics with the tools of modern linear algebra. We will call the elementary structure that establishes the species charges and a minimal set of hypothetical reactions that neutralize them a \textit{salt--charge basis}. 

It is helpful to illustrate with an example before formalizing the process for defining a salt--charge basis. Consider a solution of $\ch{Na+}$, $\ch{Cl-}$, $\ch{Mg}^{2+}$ and $\ch{SO_4^{2-}}$ in water ($\ch{H2O}$). Among these five species, one can write at most $5-1 = 4$ independent reactions to form electrically neutral entities. For instance, one could write a trivial equilibrium between the naturally uncharged water molecule and itself, and then write three equilibria that combine distinct countercharged pairs of ions to neutralize their charges:
\begin{align*}
& \ch{H2O} \leftrightharpoons \ch{H2O} \\
& \ch{Na+} + \ch{Cl-} \leftrightharpoons \ch{NaCl} \\
& \ch{Mg}^{2+} + 2 \ch{Cl-} \leftrightharpoons \ch{MgCl2} \\
& 2\ch{Na+} + \ch{SO_4^{2-}} \leftrightharpoons \ch{Na2SO4}. 
\end{align*}
\noindent Along with the list of species charges $\left\{ 0, 1, -1, 2, -2 \right\}$, these simple association equilibria define a salt--charge basis for this five-species aqueous solution, in this case involving the uncharged components $\ch{H2O}$, $\ch{NaCl}$, $\ch{MgCl2}$, and $\ch{Na2SO4}$. 

The choice of reactions to form a salt--charge basis is not generally unique, but any neutral component not defined can always be recovered by writing additional reactions involving products formed by the selected simple association equilibria. In the example system, observe that one can create the fourth possible simple binary salt, $\ch{MgSO_4}$, via the recombination reaction
\begin{equation}\label{makeMgSO4}
\ch{MgCl2} + \ch{Na2SO4} \leftrightharpoons 2 \ch{NaCl} + \ch{MgSO4} .
\end{equation}
\noindent Through this equilibrium, any thermodynamic property of $\ch{MgSO4}$ within this system---such as its chemical potential, partial molar volume, or partial molar entropy---is determined by the properties of dissolved $\ch{NaCl}$, $\ch{Na2SO4}$, and $\ch{MgCl2}$. The energetics of any multi-ion salt with higher formula-unit stoichiometry also follows from the simple association equilibria. Consider
\begin{equation*}
\ch{MgCl2} + \ch{Na2SO4} \leftrightharpoons \ch{NaCl} + \ch{Mg Na Cl SO4 },
\end{equation*}
which forms a more complex salt, $\ch{Mg Na Cl SO4 }$, from all four available ions. 

The formal procedure for creating a salt--charge basis generalizes the preceding example. Given a set of species that comprise an electrolytic solution, start by placing them in an ordered list, wherein each species' position assigns its numerical index. Let $n_\textrm{c}$ denote the number of charged species. List all $(n-n_\textrm{c})$ uncharged species first, and charged species afterward. For consistency with prior approaches, ensure that the $n$th and $(n-1)$th species have opposite charges \cite{Monroe2013,GOYAL2021137638}. Then let $\textbf{z} = [z_{1},...,z_{n}]^\top$ be a column matrix comprising the species equivalent charges in their assigned order. The ordering ensures that entries $z_1, ...,z_{n-n_\textrm{c}}$ of $\textbf{z}$ equal zero, entries $z_{n-n_\textrm{c}+1},...,z_n$ of $\textbf{z}$ are nonzero, and $z_n / z_{n-1} < 0$. 

Next, write a set of $n-1$ simple association equilibria involving the species, beginning with $n-n_\textrm{c}$ trivial reactions for the naturally uncharged species, and following with $n_\textrm{c}-1$ linearly independent association reactions, within which countercharged pairs of charged species combine to form simple binary salts. (Here `simple' means that the reactant stoichiometric coefficients are coprime positive integers.)
Let column matrix $\boldsymbol{\nu}_{i} = [\nu_{i1}, \nu_{i2},...,\nu_{in}]^\top$ contain the stoichiometric coefficients $\nu_{ij}$ of reactant species $j$ in the $i$th simple association equilibrium. For uncharged species $k=1,...,n-n_\textrm{c}$, stoichiometric column $\boldsymbol{\nu}_{k}$ will be the $k$th column of the identity matrix, notated henceforth as $\textbf{i}_k$. For $k=n-n_\textrm{c}+1,...,n-1$, $\boldsymbol{\nu}_{k}$ has two positive-integer entries, corresponding to the stoichiometric coefficients of species in a formula unit of the simple binary salt they form. The restriction to simple salts ensures that the total number of ions in each formula unit is minimal.

Guggenheim's condition for association equilibria requires that the ion stoichiometry within a formula unit of salt always serves to neutralize its charge \cite{Guggenheimbook}. In terms of reactant stoichiometry in the $i$th simple association equilibrium, the condition is expressed as 
\begin{equation}\label{Guggenheim.condition}
\boldsymbol{\nu}_{i}^\top \textbf{z} = 0,
\end{equation}
where superscript $\top$ indicates the matrix transpose. Note that this condition also holds for every trivial simple association equilibrium based on an uncharged reactant. 

It is useful to view equation \eqref{Guggenheim.condition} as a statement about a matrix inner product, namely: the Guggenheim condition requires that each column of stoichiometric coefficients $\boldsymbol{\nu}_i$ is orthogonal to the charges $\textbf{z}$. Since the set of stoichiometric columns is also linearly independent by construction, the reactant stoichiometries from the $n-1$ simple association equilibria establish an $(n-1)$-dimensional linear space orthogonal to the span of $\textbf{z}$. Moreover, the  $n$-tuple $\left\{ \boldsymbol{\nu}_1,...,\boldsymbol{\nu}_{n-1},\textbf{z} \right\}$ is linearly independent, and consequently forms a basis for the entire $n$-dimensional composition space. This set of column matrices formally represents the chosen salt--charge basis.

\section{Chemical potential in a salt--charge basis}

Selecting a salt--charge basis enables one to construct a linear transformation that maps the $n_\textrm{c}$ charged-species electrochemical potentials into $n_\textrm{c}-1$ chemical potentials of neutral salts, leaving a unique additional combination of electrochemical potentials that can be used to describe charge. The transformation is represented by an $n \times n$ matrix $\textbf{Z}$, 
\begin{equation}\label{Zdef}
\textbf{Z} = \left[ \begin{array}{c}
\boldsymbol{\nu}_{1}^\top  \\ \vdots \\ \boldsymbol{\nu}_{n-1}^\top \\ {\textbf{z}^\top}/{\|\textbf{z}\|} 
\end{array}
\right],
\end{equation}
in which $\|\textbf{v}\|= \sqrt{\textbf{v}^\top \textbf{v}}$ for any column $\textbf{v}$. Its rows are linearly independent, so $\textbf{Z}$ is invertible. Each of the first $n-1$ rows is orthogonal to the last ($n$th) row, and each of the first $n-n_\textrm{c}$ rows, which correspond to uncharged species, is orthogonal to every other row. Bear in mind that internal rows corresponding to salts, that is, rows $i = n-n_\textrm{c}+1,..., n-1$ of $\textbf{Z}$, are not necessarily mutually orthogonal. 

The transformation $\textbf{Z}$ combines species electrochemical potentials into component chemical potentials as follows. Let the column matrix $\boldsymbol{\mu} = [\mu_{1},...,\mu_{n}]^\top$ contain the species electrochemical potentials in order. Then multiplication by $\textbf{Z}$ maps $\boldsymbol{\mu}$ into new coordinates $\boldsymbol{\mu}_Z$ that stand over the chosen salt--charge basis. In the form of a matrix equation, $\boldsymbol{\mu}_Z$ is defined as
\begin{equation}\label{muZofmu}
\boldsymbol{\mu}_Z = \textbf{Z} \boldsymbol{\mu}.
\end{equation}
\noindent The first $n-1$ rows of $\boldsymbol{\mu}_Z$ so formed comprise the chemical potentials of a minimal number of neutral components, with the $i$th row representing the chemical potential of the product formed by the $i$th simple association equilibrium used to define the salt--charge basis. The last row quantifies all of the information in the set of species electrochemical potentials associated with reversible processes that do not conserve charge. It is convenient to subdivide $\boldsymbol{\mu}_Z$, as 
\begin{equation}\label{muZpartition}
\boldsymbol{\mu}_Z = \left[ \begin{array}{c}
\boldsymbol{\mu}_\nu \\
\mu_{z}
\end{array}
\right],
\end{equation} 
where the ($n-1$)-dimensional column $\boldsymbol{\mu}_\nu$ contains the $n-1$ component chemical potentials. More explicitly, if $\boldsymbol{\nu}_k$ has two nonzero entries, $\mu_{\nu,k} = ( \boldsymbol{\mu}_\nu )_k = \boldsymbol{\nu}_k^\top \boldsymbol{\mu}$ is a salt chemical potential formed by the same process used by Newman and Thomas--Alyea \cite{newman2004electrochemical}; otherwise (if $\boldsymbol{\nu}_k = \textbf{i}_k$), $\mu_{\nu,k}$ reproduces the chemical potential expected for a naturally uncharged species. As we shall see later, the remaining quantity, $\mu_{z}$, leads naturally to a definition of solution voltage. 

\section{Concentration and flux in a salt--charge basis}

The transformation to a salt--charge basis embodied by $\textbf{Z}$ can be exploited to form quantities which describe the amounts and motion of uncharged components. Natural definitions of component concentrations and component fluxes arise heuristically from an invariance argument. 

We assert as a physical principle that the elementary laws from thermodynamics and irreversible thermodynamics should retain their structures when expressed in terms of components, rather than species. The essential thermodynamic relation among species is the Euler equation, 
\begin{equation}\label{basicEulerG}
\tilde{G} = \sum_{i=1}^n \mu_i c_i ,
\end{equation}
which defines volumetric Gibbs free energy $\tilde{G}$ in terms of species molarities. The essential relation from irreversible thermodynamics is the dissipation function, written at a point within an isothermal, isobaric single phase as
\begin{equation}\label{dissipation.notmatrix}
T\dot{s} = \sum_{i=1}^n \vec{N}_i \cdot \left( - \nabla \mu_i \right),
\end{equation}
which relates local energy loss $T \dot{s}$ to the species fluxes. 

To identify a natural set of composition variables over a given salt--charge basis, require $\tilde{G}$ to retain its structure whether written in terms of species or components. Then the matrix relationship
\begin{equation}\label{EulerG}
\tilde{G} = \boldsymbol{\mu}^\top \textbf{c} = \boldsymbol{\mu}_Z^\top \textbf{c}_Z
\end{equation}
expresses $\tilde {G}$ either in terms of the species concentrations $\textbf{c}$ or a column matrix $\textbf{c}_Z$ that lists component concentrations in the chosen salt--charge basis. Since $\textbf{Z}$ is nonsingular, it follows from the second equality in equation \eqref{EulerG} and equation \eqref{muZofmu} that 
\begin{equation}\label{cZofc}
\textbf{c}_Z = \textbf{Z}^{- \top} \textbf{c},
\end{equation}
where superscript ${- \top}$ indicates the inverse transpose. 

Introducing Faraday's law helps to show how the column matrix $\textbf{c}_Z$ parses electrical information. Faraday's law for charge combines with equation \eqref{cZofc} to show that excess charge density can be expressed in two equivalent ways,
\begin{equation}\label{rhoeofcZ}
\rho_\textrm{e} = F\textbf{z}^\top \textbf{c} ~~\textrm{or}~~ \rho_\textrm{e} = F \left( \textbf{Z} \textbf{z} \right)^\top  \textbf{c}_Z ,
\end{equation}
in which the distributive property of the matrix transpose was used to rearrange the last equality. Since the first $n-1$ rows of $\textbf{Z}$ are orthogonal to $\textbf{z}^\top$, $\textbf{Z} \textbf{z} = \left\| \textbf{z} \right\|  \textbf{i}_n$ by equation \eqref{Zdef}, and consequently equation \eqref{rhoeofcZ} implies that
\begin{equation}
\left( \textbf{c}_Z \right)_n = \frac{\rho_\textrm{e}}{F \left\| \textbf{z} \right\|},
\end{equation}
i.e., the last entry of $\textbf{c}_Z$ quantifies excess charge density. Thus $\textbf{c}_Z$ naturally partitions into a column that represents an ordered set of $n-1$ neutral \textit{component concentrations}, written as $\textbf{c}_\nu$, followed by an entry expressing the excess charge in molar units,
\begin{equation}\label{cZpartition}
\textbf{c}_Z = \left[\begin{array}{c}
\textbf{c}_\nu \\
\dfrac{\rho_\textrm{e}}{F \left\| \textbf{z} \right\|}
\end{array}
\right],
\end{equation}
which is zero when the local electroneutrality approximation holds. Entry $c_{\nu,i} = \left( \textbf{c}_\nu \right)_i$ of the component concentrations $\textbf{c}_\nu$ represents the molarity of the product formed by the $i$th simple association equilibrium. 

Natural flux descriptors over a salt--charge basis can be created by a similar route, by demanding that the structure of the energy dissipation $T\dot{s}$ is preserved. Assemble the species fluxes into a column matrix $\vec{\textbf{n}} = [ \vec{N}_1,...,\vec{N}_n ]^\top$.\footnote{The overarrow indicates that each entry in the column $\vec{\textbf{n}}$ is a vector quantity. Whenever the matrix multiplication makes sense, $\textbf{A} \vec{\textbf{B}}$ forms a matrix $\vec{\textbf{C}}$ with vector entries $\vec{C}_{ij} = \sum_{k} A_{ik} \vec{B}_{kj}$, and $\vec{\textbf{A}} \cdot \vec{\textbf{B}}$ forms $\textbf{C}$, with scalar entries $C_{ij} = \sum_{k} \vec{A}_{ik} \cdot \vec{B}_{kj}$.} Structure preservation of equation \eqref{dissipation.notmatrix} requires that
\begin{equation}\label{dissipation}
T \dot{s} = \vec{\textbf{n}}^\top \cdot \left( - \nabla \boldsymbol{\mu} \right) = \vec{\textbf{n}}_Z^\top \cdot \left( - \nabla \boldsymbol{\mu}_Z \right),
\end{equation}
where the column $\vec{\textbf{n}}_Z$ lists fluxes over the chosen salt--charge basis $\textbf{Z}$. (The notation $\nabla \textbf{v}$ indicates a column matrix whose $i$th entry is the vector $\nabla v_i$, where $v_i$ is the $i$th entry of $\textbf{v}$.)  The transformation from $\vec{\textbf{n}}$ into $\vec{\textbf{n}}_Z$ therefore must be
\begin{equation}\label{nZofn}
\vec{\textbf{n}}_Z = \textbf{Z}^{-\top} \vec{\textbf{n}},
\end{equation}
identical to the mapping that sends species concentrations into component concentrations. 

Faraday's law---this time, for current---also provides detail about the new fluxes $\vec{\textbf{n}}_Z$. In matrix form, Faraday's law casts the current density $\vec{i}$ in terms of species fluxes as
\begin{equation}\label{FaradayForCurrent}
\vec{i} = F \textbf{z}^\top \vec{\textbf{n}} = F \left( \textbf{Z} \textbf{z} \right)^\top \vec{\textbf{n}}_Z.
\end{equation}
\noindent Again, the fact that $\textbf{Z} \textbf{z} = \left\| \textbf{z} \right\| \textbf{i}_n$ comes into play, yielding
\begin{equation}
\left( \vec{\textbf{n}}_Z \right)_n = \frac{\vec{i}}{F\left\| \textbf{z} \right\|}.
\end{equation}
\noindent Thus the last entry of $\vec{\textbf{n}}_Z$ represents a renormalized current density. It is again natural to partition $\vec{\textbf{n}}_Z$, as
\begin{equation}\label{nZpartition}
\vec{\textbf{n}}_Z = \left[ \begin{array}{c}
\vec{\textbf{n}}_\nu \\
\dfrac{\vec{i}}{F\left\| \textbf{z} \right\|}
\end{array}
\right],
\end{equation}
in which $\vec{\textbf{n}}_\nu$ is an $(n-1)$-dimensional column matrix representing the set of \emph{total component fluxes}. 

Other fundamental thermodynamic laws also readily transform. For example, prior definitions imply that
\begin{equation}\label{GibbsDuhem0}
\sum_{i=1}^n c_i \nabla \mu_i = \textbf{c}^\top \nabla \boldsymbol{\mu} 
= \textbf{c}_Z^\top \nabla \boldsymbol{\mu}_Z = \vec{0},
\end{equation}
so the change to a salt--charge basis retains a structurally identical isothermal, isobaric Gibbs--Duhem equation.

One can also cast the total species concentration $c_\textrm{T}$ in terms of the component concentrations. Letting $\textbf{1}$ stand for a column matrix whose entries are all `1', one has that
\begin{equation}\label{cTdef}
c_\textrm{T} = \sum_{i=1}^n c_i  = \textbf{1}^\top \textbf{c} =  \boldsymbol{\nu}_Z^{\top} \textbf{c}_Z, 
\end{equation}
in which the column matrix $\boldsymbol{\nu}_Z$, defined as
\begin{equation}\label{nuZdef}
\boldsymbol{\nu}_Z = \textbf{Z} \textbf{1}  = \left[ \begin{array}{c} 
\boldsymbol{\nu} \\
\dfrac{\textbf{z}^\top \textbf{1} }{ \left\| \textbf{z} \right\|}
\end{array}\right],
\end{equation}
summarizes information about net reactant stoichiometry in the simple association equilibria. The second equality here indicates that the stoichiometric column $\boldsymbol{\nu}_Z$ also partitions into $(n-1)$ entries $\boldsymbol{\nu}$ that describe components and a final entry associated with charge. When component $i < n$ derives from a naturally uncharged species, the corresponding entry of $\boldsymbol{\nu}$, that is, $\nu_{i}$, is 1; when component $i<n$ is a salt, $\nu_{i}$ is the total number of ions in its formula unit.

Molar concentrations are convenient for stating the thermodynamic laws, but they make characterization efforts error-prone because molarity varies with temperature and pressure at fixed composition. It is useful to introduce two alternative composition descriptors, called the \emph{species fractions} $\textbf{y}$ and \emph{component fractions} $\textbf{y}_Z$, and defined as
\begin{equation}\label{componentfractionsdef}
\textbf{y} = \frac{1}{c_\textrm{T}} \textbf{c} ~~~~\textrm{and}~~~~\textbf{y}_Z = \frac{1}{c_\textrm{T}} \textbf{c}_Z,
\end{equation}
respectively. These afford the fundamental properties that
\begin{equation}\label{ysumandyZsum}
\textbf{1}^\top \textbf{y} = 1 ~~~~\textrm{and}~~~~\boldsymbol{\nu}_Z^\top \textbf{y}_Z = 1,
\end{equation}
and are independent of temperature or pressure.

\section{Governing equations over a salt--charge basis}\label{sec:GEs_saltcharge}

The move to a salt--charge basis preserves the structure of material balances. In the absence of homogenous reactions, all the species continuity equations are summarized in matrix form as
\begin{equation}\label{speciescontinuities}
\frac{\partial \textbf{c}}{\partial t} = - \nabla \cdot \vec{\textbf{n}}.
\end{equation}
\noindent (Given any column $\vec{\textbf{v}}$, the operation $\nabla \cdot \vec{\textbf{v}}$ forms a column matrix whose $i$th entry is the scalar $\nabla \cdot \vec{v}_i$.) Multiplication by $\textbf{Z}^{-\top}$ and use of equations \eqref{cZofc} and \eqref{nZofn} yields
\begin{equation}\label{cZbalance}
\frac{\partial \textbf{c}_Z}{\partial t} = - \nabla \cdot \vec{\textbf{n}}_Z,
\end{equation}
confirming structure preservation. Bringing in the decompositions from equations \eqref{cZpartition} and \eqref{nZpartition}, one can alternatively express equation \eqref{cZbalance} as a set of $n-1$ component mole balances,
\begin{equation}\label{cnubalances}
\frac{\partial \textbf{c}_\nu}{\partial t} = - \nabla \cdot \vec{\textbf{n}}_\nu,
\end{equation}
along with the single equation
 \begin{equation}\label{chgcont}
 \frac{\partial \rho_\textrm{e}}{\partial t} = - \nabla \cdot \vec{i} ,
 \end{equation}
which establishes charge continuity. Note that any model governed by species material balances which also incorporates Faraday's law is compatible with Maxwellian electrodynamics, because Maxwell's equations imply a charge-continuity relationship identical to equation \eqref{chgcont}. 

Because the transformations that form $\textbf{c}_Z$ and $\vec{\textbf{n}}_Z$ preserve the structure of the dissipation function, they also preserve the structure of the Onsager--Stefan--Maxwell constitutive laws. Equation \eqref{OSMequation} takes the matrix form
\begin{equation}\label{OSMmatrix}
- \nabla \boldsymbol{\mu} = \textbf{M} \vec{\textbf{n}}.
\end{equation}
\noindent Equations \eqref{muZofmu} and \eqref{nZofn} show that this transforms to
\begin{equation}\label{OSM_saltcharge}
- \nabla \boldsymbol{\mu}_Z  = \textbf{M}_Z \vec{\textbf{n}}_Z 
\end{equation}
over a salt--charge basis. The transport matrix $\textbf{M}_Z$ in the salt--charge representation of the Onsager--Stefan--Maxwell laws relates to the matrix $\textbf{M}$ from equation \eqref{transport matrix} through 
\begin{equation}\label{MZofM}
\textbf{M}_Z = \textbf{Z} \textbf{M} \textbf{Z}^{\top},
\end{equation}
a congruence transformation. It follows from this congruence that $\textbf{M}$ is symmetric if and only if $\textbf{M}_Z$ is symmetric, so that the Onsager--Stefan--Maxwell transport coefficients retain symmetry when expressed over a salt--charge basis. Sylvester's law of inertia implies further that the signatures and ranks of $\textbf{M}$ and $\textbf{M}_Z$ equate. As mentioned before, the non-dissipative nature of convection implies that $\textbf{M}$ affords $\textbf{c}$ as its sole null eigenvector; in light of congruence relation \eqref{MZofM} and Sylvester's law, this implies that $\textbf{c}_Z$ is the sole null eigenvector of $\textbf{M}_Z$. 

Thus, transformation to a salt--charge basis preserves the structure of the differential balance equations as well as all spectral properties of the Onsager--Stefan--Maxwell equations. Both $\textbf{M}$ and $\textbf{M}_Z$ are symmetric matrices, and each affords a single null eigenvector. Although the null spaces of $\textbf{M}$ and $\textbf{M}_Z$ generally differ, each matrix affords just one null eigenvector ($\textbf{c}$ and $\textbf{c}_Z$, respectively). The nonzero eigenvalues of $\textbf{M}$ generally differ in magnitude from those of $\textbf{M}_Z$, but both sets are all positive.

\section{The salt--charge potential}

It is prudent to introduce a thermodynamic potential with units of voltage to quantify electrical energy and power. To put electricity in a thermodynamic context, however, one must acknowledge an essential limitation on how electrochemical equilibria are parametrized. 

When proposing the electrochemical potential concept, Guggenheim introduced constitutive laws
\begin{equation}\label{mui_Gugg}
\mu_i = RT \ln a_i + F z_i \Phi,
\end{equation}
in which $a_i$ represents the chemical activity of species $i$ and $\Phi$ is an electric potential in the solution phase \cite{Guggenheim,Guggenheimbook,newman2004electrochemical}. Observing that the equilibrium compositions of practical macroscopic systems cannot be adjusted along paths that violate global electroneutrality, Guggenheim noted that the partitioning of electrochemical potentials into chemical and electrical parts is necessarily ambiguous. Therefore a model of electrolyte energetics must never depend on the chemical activity of a single charged species in isolation \cite{Guggenheimbook}. 
To cast this notion formally, define a non-neutral free-energy contribution $\mu_{z}$, written in terms of Guggenheim's species activities and solution potential as
\begin{equation}\label{murhoedef}
\mu_{z} = \frac{1}{\left\| \textbf{z} \right\|} \sum \limits_{i=1}^n z_i \mu_i  = \frac{RT}{\left\| \textbf{z} \right\|} \sum \limits_{i=1}^n z_i \ln a_i + F \left\| \textbf{z} \right\| \Phi. 
\end{equation}
\noindent Guggenheim's principle states that no macroscopic experiment can discern the term that involves the solution potential $\Phi$ in this expression from those involving the species activities $a_i$ \cite{Guggenheim}. 

Rather than using the solution potential $\Phi$, we instead employ a quantity we call the \emph{salt--charge potential} $\Phi_z$, defined as
\begin{equation}\label{Phi_nu}
\Phi_z = \frac{\mu_{z}}{F \left\| \textbf{z} \right\| }.
\end{equation}
Guggenheim's principle requires that no experiment can distinguish chemical and electrical contributions to $\Phi_z$. Despite this ambiguity, one can still understand the salt--charge potential $\Phi_z$ as a type of electrical potential. Inserting decompositions \eqref{muZpartition} and \eqref{cZpartition}, then introducing definition \eqref{Phi_nu} recasts the Euler equation for $\tilde G$ as
\begin{equation}\label{EulerGpartition}
\tilde{G} = \boldsymbol{\mu}_\nu^\top \textbf{c}_\nu +  \Phi_z \rho_{\textrm{e}}.
\end{equation}
\noindent Here, the $\Phi_z \rho_\textrm{e}$ term apparently quantifies excess coulomb energy, that is, the available energy associated with imbalanced charge. Similarly, inserting decompositions \eqref{muZpartition} and \eqref{nZpartition} restates the dissipation as
\begin{equation}\label{dissipationpartition}
T \dot{s} = \vec{\textbf{n}}_\nu^\top \cdot \left( - \nabla \boldsymbol{\mu}_\nu \right) + \vec{i} \cdot \left( - \nabla \Phi_z \right).
\end{equation}
\noindent The $- \vec{i} \cdot \nabla \Phi_z$ term is the local ohmic loss---the rate of energy dissipation associated with electrical current flow. Observe that local electroneutrality makes the coulomb term disappear from $\tilde{G}$, but leaves the ohmic term in $T \dot{s}$. 

Although the species material balances summarized in equation \eqref{speciescontinuities} imply charge-continuity equation \eqref{chgcont}, and are therefore compatible with Maxwell's equations, equilibrium thermodynamics cannot provide insight into the Maxwellian electric field, because Guggenheim's principle mandates that it is impossible to determine experimentally whether or not $-\nabla \Phi_z$ is a purely electrical quantity. 

Still, the energy density and dissipative power associated with salt--charge potential (and its gradient) have  unambiguous meanings, because the coulomb term in $\tilde{G}$ and ohmic term in $T \dot{s}$ do not vary with the choice of components. Whereas one can only speak of `a' salt--charge basis since $\textbf{Z}$ is not always unique, one may refer to `the' salt--charge potential, which is unique for a given $\textbf{z}$.

\section{Thermodynamic factors in a salt--charge basis}\label{sec:DarkenZ}
To implement a simulation of any transport system, one requires constitutive relationships that express how thermodynamic potentials depend on the concentrations of extensive properties. The particular mapping that sends composition gradients into electrochemical-potential gradients in isothermal, isobaric situations is expressed by a matrix of equilibrium properties commonly referred to as \emph{thermodynamic factors}. 

For an isothermal, isobaric, $n$-species mass-transport system, the thermodynamic factors depend parametrically on $\tfrac{1}{2}n\left( n-1 \right)$ composition-dependent state functions---one state function for every pair of species \cite{Monroe2015,GOYAL2021137638,VanbruntAIChE}. In their discussion of Onsager reciprocal relations for multicomponent diffusion, Monroe and Newman introduced an $\left( n-1 \right) \times \left( n- 1 \right)$ matrix of dimensionless Darken factors $\textbf{Q}$ \cite{Monroe2006}, defined such that
\begin{equation}\label{dmuofQ}
\diag \left( \textbf{y} \right) \nabla \boldsymbol{\mu} = RT \left[ 
\begin{array}{c} 
\textbf{I}_{n-1} \\
-\textbf{1}^\top 
\end{array}
\right] \textbf{Q} \left[ \begin{array}{cc}
\textbf{I}_{n-1} & \textbf{o} 
\end{array}
\right]^\top \nabla \textbf{y},
\end{equation}
where $\textbf{I}_{n-1}$ indicates the $\left( n -1 \right) \times \left( n-1 \right)$ identity matrix, $\textbf{o}$ represents a column of zeroes, and the linear operator $\diag\left( \textbf{v} \right)$ forms column matrix $\textbf{v}$ into a diagonal square matrix whose $i$th diagonal entry is $v_i$.\footnote{
Generally $\diag \left( \textbf{a} \right) \textbf{b} = \diag \left( \textbf{b} \right) \textbf{a}$; both products form a column $\textbf{c}$ with entries $c_i = a_i b_i$. Also $\textbf{1}^\top \diag(\textbf{a}) = \textbf{a}^\top$ and $\diag(\textbf{a}) \textbf{1} = \textbf{a}$.
The notation $\diag \left( \textbf{a} \right)^{-1}$ indicates the diagonal matrix whose $i$th entry is $1/a_i$, such that $\diag \left( \textbf{a} \right)^{-1} \textbf{a} = \textbf{1}$. 
}
Each entry $Q_{ij}$ of $\textbf{Q}$ represents the thermodynamic derivative of the (log) activity of species $i$ with respect to the particle fraction of species $j < n$, leaving temperature, pressure, and all particle fractions save those of species $j$ and $n$ fixed. (Experiments to measure $Q_{ij}$ adopt a convention that when one varies $y_j$, one specifically adjusts $y_n$ to maintain the constraint that $\textbf{1}^\top \textbf{y} = 1$.) In terms of Guggenheim's species activities, $Q_{ij}$ is defined as
\begin{align}\label{Qijdef}
Q_{ij} = y_i \left( \frac{\partial \ln a_i}{\partial y_j} \right)_{T,p,y_{k \ne j, n}} = \delta_{ij} + y_i \left( \frac{\partial \ln \lambda_i}{\partial y_j} \right)_{T,p,y_{k \ne j, n}}, 
\end{align}
in which $\delta_{ij}$ represents the Kronecker delta. The expression on the right of equation \eqref{Qijdef} introduces the \emph{activity coefficient} of species $i$, $\lambda_i$, defined such that $a_i = \lambda_i y_i$. Introducing $\lambda_i$ emphasizes the expectation that $\textbf{Q} = \textbf{I}_{n-1}$ within solutions whose mixing free energies are thermodynamically ideal. 

Although the simplicity of $\textbf{Q}$ in the ideal case is useful, Maxwell relations among these Darken factors in non-ideal situations are obscure. It is sometimes convenient to work instead with the $(n-1) \times (n-1)$ Hessian matrix $\textbf{K}$, 
\begin{equation}
K_{ij} = \frac{\partial^2}{\partial y_i \partial y_j} \left( \frac{\tilde G}{RT c_\textrm{T} } \right)_{T,p,y_{k \ne i,j,n}},
\end{equation}
for which Maxwell relations imply that $\textbf{K} = \textbf{K}^\top$ \cite{Monroe2015}. Thermodynamic stability demands that $\textbf{K}$ is positive semidefinite \cite{GOYAL2021137638}. The Darken factors $\textbf{Q}$ depend on $\textbf{K}$ through
\begin{equation}\label{KofYinvQ}
\textbf{Q} = \textbf{Y}^{-1} \textbf{K}, 
\end{equation}
where the definition
\begin{equation}\label{Yinvdef}
\textbf{Y}^{-1} = \left[ \begin{array}{cc}
\textbf{I}_{n-1} & \textbf{o} 
\end{array} \right]
\left( \diag \left( \textbf{y} \right) - \textbf{y} \textbf{y}^\top \right) \left[ \begin{array}{c}
\textbf{I}_{n-1} \\ 
\textbf{o}^\top 
\end{array}
\right] 
\end{equation}
puts in matrix form the $(n-1) \times (n-1)$ inverse composition matrix $\textbf{Y}^{-1}$ employed by Monroe and Newman \cite{Monroe2009}. Substituting equations \eqref{KofYinvQ} and \eqref{Yinvdef} into equation \eqref{dmuofQ} and algebraically simplifying the result gives
\begin{equation}\label{dmuofK}
\nabla \boldsymbol{\mu} = RT \left( \textbf{I} - \textbf{1} \textbf{y}^\top \right) \left[ \begin{array}{cc}
\textbf{K} & \textbf{o}  \\
\textbf{o}^\top & 0 
\end{array} \right] \nabla \textbf{y} ,
\end{equation}
which shows how the isothermal, isobaric composition dependence of $\nabla \boldsymbol{\mu}$ is parametrized by the $\tfrac{1}{2} n \left( n-1\right)$ independent entries of $\textbf{K}$. 
Observe that the block matrix here has a nullspace spanned by the null eigenvector $\textbf{i}_n$, which reflects the fact that gradients $\nabla y_n$ have no independent effect on species electrochemical potentials.

Some of the benefits of the Darken matrix $\textbf{Q}$ can be brought into the analysis of equations expressed in terms of $\textbf{K}$. The composition Hessian separates into two parts, 
\begin{equation}
\textbf{K} = \textbf{Y} + \Delta \textbf{K},
\end{equation}
such that $\textbf{Y}$ accounts for derivatives of ideal mixing free energy and $\Delta \textbf{K}$ expresses the Hessian of excess free energy. The composition matrix $\textbf{Y}$ here is written explicitly as
\begin{equation}\label{Ydef}
\textbf{Y} = \left[ \begin{array}{cc}
\textbf{I}_{n-1} & \textbf{o} 
\end{array} \right]
\left( \diag \left( \textbf{y} \right)^{-1} + \frac{\textbf{1} \textbf{1}^\top }{\textbf{i}_n^\top \textbf{y}}  \right) \left[ \begin{array}{c}
\textbf{I}_{n-1} \\ 
\textbf{o}^\top 
\end{array}
\right].
\end{equation}
\noindent Upon substitution into equation \eqref{dmuofK}, algebraic simplification yields an equivalent alternative form,
\begin{align}\label{dmuwithdeltaK}
\nabla \boldsymbol{\mu} = \, &RT \diag \left( \textbf{y} \right)^{-1} \nabla \textbf{y} \nonumber \\ & + RT \left( \textbf{I} - \textbf{1} \textbf{y}^\top \right) \left[ \begin{array}{cc} 
\Delta \textbf{K} & \textbf{o} \\
\textbf{o}^\top & 0 
\end{array} \right] \nabla \textbf{y},
\end{align}
in which the $(n-1) \times (n-1)$ matrix $\Delta \textbf{K}$ is symmetric, and vanishes when mixing is ideal.

The nonideal term in equation \eqref{dmuwithdeltaK} translates readily into a salt--charge basis. Use the congruence relationship 
\begin{equation}\label{KZdef}
\Delta \textbf{K}_Z = \textbf{Z} \left[ \begin{array}{cc}
\Delta \textbf{K} & \textbf{o}  \\
\textbf{o}^\top & 0 
\end{array} \right] \textbf{Z}^\top
\end{equation}
to define a transformed excess Hessian $\Delta \textbf{K}_Z$. All $n^2$ entries of this symmetric matrix are generally nonzero. The nullspace of $\Delta \textbf{K}_Z$ has a minimum dimension of $1$, however, because $\textbf{Z}^{-\top} \textbf{i}_n$ is always a null eigenvector. Multiplying equation \eqref{dmuwithdeltaK} through by $\textbf{Z}$ and inserting $\Delta \textbf{K}_Z$, one finds
\begin{align}\label{dmuZexpansion}
\nabla \boldsymbol{\mu}_Z = &RT \textbf{Z} \diag \left( \textbf{y} \right)^{-1} \nabla \textbf{y} \nonumber \\ &+ RT \left( \textbf{I} - \boldsymbol{\nu}_Z \textbf{y}_Z^\top \right) \Delta \textbf{K}_Z \nabla \textbf{y}_Z
\end{align} 
after using equations \eqref{muZofmu}, \eqref{cZofc}, and \eqref{ysumandyZsum} to simplify. The assertion that species potentials within an electrolyte at constant temperature and pressure depend on all the independent species fractions necessitates that the differential component potentials have this structure. 

It is noteworthy that the ideal part of the component chemical-potential gradients---the first term on the right of equation \eqref{dmuZexpansion}---involves entries of the salt--charge transformation $\textbf{Z}$ as prefactors. Thus the ideal mixing free energy of a dissolved salt scales linearly with its ion stoichiometry, in addition to the ideal logarithmic dependence on its mole fraction. Note also that multiplication from the left by $\textbf{c}_Z^\top = c_\textrm{T} \textbf{y}_Z^\top$ verifies that the Gibbs--Duhem equation is satisfied by constitutive laws \eqref{dmuZexpansion} for any choice of the symmetric excess Hessian matrix $\Delta \textbf{K}_Z$. 

Although equation \eqref{dmuZexpansion} is formally correct, the parameters within $\Delta \textbf{K}_Z$ are not all measurable. In particular, as discussed after equation \eqref{murhoedef}, adherence to Guggenheim's principle demands that the coefficients of $\nabla \textbf{y}_Z$ on the right side of the equation for $\nabla \mu_{z}$ should remain ambiguous because they cannot be disentangled with equilibrium experiments. Therefore it is appropriate to discard this component of the expansion in equation \eqref{dmuZexpansion}. 

The notation is simplified by defining an $n \times (n-1)$ matrix $\textbf{N}$, whose columns represent the reactant stoichiometry in the simple association equilibria: 
\begin{equation}\label{Nmatdef}
\textbf{N} = \left[ \begin{array}{ccc}
\boldsymbol{\nu}_{1} & \dots & \boldsymbol{\nu}_{n-1}
\end{array}
\right], ~~~\textrm{so}~~~\textbf{Z} = \left[ \begin{array}{c}
\textbf{N}^\top \\
\dfrac{\textbf{z}^\top}{\left\| \textbf{z} \right\| }
\end{array}
\right].
\end{equation} 
\noindent Further observing that $\boldsymbol{\nu}=\textbf{N}^\top \textbf{1}$, the general expression
\begin{align}\label{gradmunu}
\nabla \boldsymbol{\mu}_\nu = &RT \textbf{N}^\top \diag \left( \textbf{y} \right)^{-1} \nabla \textbf{y} \nonumber \\
&+ RT \left( \left[ \begin{array}{cc} 
\textbf{I}_{n-1} & \textbf{o} 
\end{array} \right] - \boldsymbol{\nu} \textbf{y}_Z^\top \right) \Delta \textbf{K}_Z \nabla \textbf{y}_Z
\end{align} 
expresses the portion of the expansion that remains after striking the last row of matrix equation \eqref{dmuZexpansion} to accommodate Guggenheim's principle. In the next section, this form of the component chemical-potential gradients will facilitate analyzing constraints on properties under the electroneutrality approximation. 

\section{Structural implications of electroneutrality}
Imposing local electroneutrality on the Onsager--Stefan--Maxwell transport model impacts both the variables involved and the parameters that quantify material properties. The approximation separately impacts the thermodynamic and dynamical aspects of the general theory.

\subsection{Electroneutral composition}\label{subsec:ElectroneutComp}
The simplest consequence of electroneutrality is that it limits the available composition space, because one charged-species concentration becomes linearly dependent on the others. It is useful to establish some relationships that demonstrate the effect of this constraint.

When excess charge density vanishes everywhere, the $n$-dimensional column of species concentrations $\textbf{c}$ becomes limited to a set of locally electroneutral species concentrations, which we notate as $\textbf{c}^0$. Necessarily, any solution compositions in the space that $\textbf{c}^0$ occupies must be instantiable by assembling uncharged components in the appropriate proportions and then dissociating them into their constituent species. This idea can be put in mathematical terms by leveraging a salt--charge basis. Over such a basis $\textbf{c}^0$ is determined by the mapping
\begin{equation}\label{c0def}
\textbf{c}^0 = \textbf{Z}^\top \left[ \begin{array}{c}
\textbf{c}_\nu \\
0
\end{array} \right] = \textbf{N} \textbf{c}_\nu,
\end{equation}
which results from equations \eqref{cZofc} and \eqref{cZpartition} when $\rho_\textrm{e} = 0$. Thus the first $(n-1)$ columns of $\textbf{Z}^\top$ (that is, the columns of $\textbf{N}$) serve as a basis for the subspace occupied by $\textbf{c}^0$, and the $(n-1)$-dimensional column of electroneutral component concentrations $\textbf{c}_\nu$ expresses the coordinates of a given $\textbf{c}^0$ over this basis. Electroneutrality constrains the total concentration's domain in a similar way. The electroneutral total concentration, $c_\textrm{T}^0$, depends on $\textbf{c}_\nu$ as
\begin{equation}\label{cT0def}
c_\textrm{T}^0 = \textbf{1}^\top \textbf{c}^0 = 
\boldsymbol{\nu}_Z^\top \left[ \begin{array}{c}
\textbf{c}_\nu \\
0
\end{array} \right] = \boldsymbol{\nu}^\top \textbf{c}_\nu , 
\end{equation}
simplified with the definition of $\boldsymbol{\nu}$ from equation \eqref{nuZdef}.

Some additional physical principles outside thermodynamics restrict the range of neutral compositions. Since negative species molarities are meaningless, every entry of $\textbf{c}^0$ must be non-negative. Because mass transport cannot occur in the absence of species, $c_\textrm{T}^0$ is strictly positive. These inequalities place bounds on the entries of the component-composition column $\textbf{c}_\nu$. 

The restrictions on $\textbf{c}^0$ and $c_\textrm{T}^0$ also imply that every entry of the electroneutral species fractions, $\textbf{y}^0$, defined as 
\begin{equation}\label{y0ofynu}
\textbf{y}^0 = \frac{1}{c_\textrm{T}^0} \textbf{c}^0 
= \textbf{Z}^\top \left[ \begin{array}{c}
\textbf{y}_\nu \\
0
\end{array} \right] = \textbf{N} \textbf{y}_\nu,
\end{equation}
must be strictly non-negative. Under electroneutrality this object retains the property that $\textbf{1}^\top \textbf{y}^0 = 1$, established by equation \eqref{ysumandyZsum}. Furthermore, the condition
\begin{equation}\label{ynusum}
\boldsymbol{\nu}^\top \textbf{y}_\nu = 1 ~~~~~~~~\textrm{(electroneutral)}
\end{equation}
constrains component fractions under electroneutrality.

Note that requiring $\textbf{c}^0$ or $\textbf{y}^0$ to have non-negative entries does not ensure that every component concentration within $\textbf{c}_\nu$ (or $\textbf{y}_\nu$) is non-negative. Over a salt--charge basis, negative component concentrations can occur when the choice of products in the simple association equilibria is not unique. Consider the solution of $\ch{Na+}$, $\ch{Cl-}$, $\ch{Mg}^{2+}$ and $\ch{SO_4^{2-}}$ in $\ch{H2O}$ from section \ref{sec:construction} by way of example. The electroneutral composition space for this solution reaches extremes corresponding to four distinct binary electrolytes: aqueous $\ch{NaCl}$ with a trace amount of $\ch{MgSO4}$; aqueous $\ch{MgCl2}$ with trace $\ch{Na2SO4}$; aqueous $\ch{Na2SO4}$ with trace $\ch{MgCl2}$; and aqueous $\ch{MgSO4}$ with trace $\ch{NaCl}$. Suppose the neutral salts for a salt--charge basis are chosen to be $\ch{NaCl}$, $\ch{MgCl2}$, and $\ch{Na2SO4}$, as in section \ref{sec:construction}. In the laboratory, these salts cannot be mixed with water to produce, say, 1 M aqueous $\ch{MgSO4}$. To formulate 1 M $\ch{MgSO4}$ with the given precursor salts, one would have to add $\ch{Na2SO4}$ to supply the sulfate, add $\ch{MgCl2}$ to introduce magnesium, and then utilize a separation process (e.g., precipitation, heterogeneous extraction, {etc.}) to remove $\ch{NaCl}$. Thus, aqueous 1 M $\ch{MgSO4}$ is represented by a $\textbf{c}_\nu$ column with entries for 1 M $\ch{Na2SO4}$, 1 M $\ch{MgCl2}$, and $-2$ M $\ch{NaCl}$. Although the $\ch{NaCl}$ molarity is negative, that of every ion remains non-negative. Since $\ch{MgSO4}$ can be formed by recombination reaction \eqref{makeMgSO4}, its chemical potential depends unambiguously on those of $\ch{Na2SO4}$, $\ch{MgCl2}$, and $\ch{NaCl}$. The negative concentration just indicates that $\ch{NaCl}$ is a product of a recombination reaction involving the salt--charge basis' components, rather than a reactant.

\subsection{Component activity coefficients}

Electroneutrality has ramifications for Guggenheim's constitutive framework. Definition \eqref{mui_Gugg} can be rewritten in terms of species activity coefficients $\lambda_i^0$ as
\begin{align}\label{mui0oflambda}
\mu_i^0 &= \mu_i^\ominus + RT \ln \left( \lambda_i^0 y_i^0 \right) + z_i F \Phi,~~~~\textrm{or} \nonumber \\\boldsymbol{\mu}^0 &= \boldsymbol{\mu}^\ominus + RT \ln \left( \textbf{y}^0 \right) + RT \ln \left( \boldsymbol{\lambda}^0 \right) + F \Phi \textbf{z} ,
\end{align}
where the composition-independent parameters $\mu_i^\ominus$ that make up the $n$-dimensional column $\boldsymbol{\mu}^\ominus$ quantify species electrochemical potentials in a secondary reference state, and $\ln \left( \textbf{v} \right)$ represents a column matrix wherein each entry is the natural logarithm of the corresponding entry of $\textbf{v}$. Thermodynamic consistency mandates that every activity coefficient comprising $\boldsymbol{\lambda}^0$ is positive. Since $( \textbf{c}^0 )^\top \textbf{z} = 0$ by design, these constitutive laws meet the requirement that
\begin{equation}\label{G0twoways}
\tilde{G} = \left( \textbf{c}^0 \right)^\top \boldsymbol{\mu}^0  = \textbf{c}_\nu^\top \boldsymbol{\mu}_\nu^0
\end{equation}
as a consequence of equations \eqref{muZofmu} and \eqref{c0def}. 

Constitutive laws for component chemical potentials in accord with Guggenheim's formulation are identified by using the second equality in equation \eqref{G0twoways}, which implies
\begin{equation}\label{Gimplication}
\left( \textbf{c}^0 \right)^\top \textbf{Z}^{-1} \left( \textbf{Z} \boldsymbol{\mu}^0 -  \boldsymbol{\mu}_Z^0 \right) = 0 .
\end{equation}
\noindent Insertion of equations \eqref{Nmatdef} and \eqref{mui0oflambda} into equation \eqref{Gimplication}, followed by partitioning of $\boldsymbol{\mu}_Z^0$ with equation \eqref{muZpartition}, application of the relation $\textbf{Z}\textbf{z} = \left\| \textbf{z} \right\| \textbf{i}_n$, and substitution of $\Phi$ in favor of $\Phi_z^0$ with equations \eqref{murhoedef} and \eqref{Phi_nu}, reveals that Guggenheim's laws are consistent with component chemical potentials constituted by equations of the form
\begin{equation}\label{munu0_constitutive.scalarform}
\mu_{\nu,k}^0 = \mu_{\nu,k}^\ominus + RT \sum \limits_{m=1}^n \nu_{km} \ln y_m^0 + RT \nu_k \ln \lambda_{\nu,k}^0 ,
\end{equation}
wherein $\mu_{\nu,k}^\ominus \left( T, p \right)$ is the chemical potential of component $k$ in the reference state and $\nu_k$ is the $k$th entry of the column $\boldsymbol{\nu}$ that quantifies total component stoichiometry. The parameter $\lambda_{\nu,k}^0$ is called the \emph{mean molar activity coefficient} of component $k$. 

All of the component chemical-potential constitutive laws can therefore be summarized by a matrix equation,
\begin{equation}\label{munu0_constitutive}
\boldsymbol{\mu}_\nu^0 = \boldsymbol{\mu}_\nu^\ominus + \boldsymbol{\mu}_\nu^{0,\textrm{ideal}} + RT \diag \left( \boldsymbol{\nu} \right) \ln \boldsymbol{\lambda}_\nu^0 ,
\end{equation}
in which the term 
\begin{equation}
\boldsymbol{\mu}_\nu^{0,\textrm{ideal}} = RT \textbf{N}^\top \ln \textbf{y}^0
\end{equation}
accounts for the part of a component's chemical potential that is attributable to ideal mixing of its constituent species, and the reference chemical potentials and mean molar activity coefficients of each component make up the $(n-1)$-dimensional columns $\boldsymbol{\mu}_\nu^\ominus$ and $\boldsymbol{\lambda}_\nu^0$, respectively: the expression
\begin{equation}\label{meanmolaractivitydef}
\boldsymbol{\mu}_\nu^\ominus = \textbf{N}^\top \boldsymbol{\mu}^\ominus,~~~\textrm{or}~~~\mu_{\nu,k}^\ominus = \sum_{m=1}^n \nu_{km} \mu_{m}^\ominus,
\end{equation}
where $k \in \left\{ 1, ..., n-1 \right\}$, shows how reactant stoichiometries in the simple association equilibria relate the reference potentials of components and species, and
\begin{align}\label{refchempotdef}
\ln \boldsymbol{\lambda}_\nu^0 &= \diag \left( \boldsymbol{\nu} \right)^{-1} \textbf{N}^\top \ln \boldsymbol{\lambda}^0,~~~\textrm{or} \nonumber \\ 
\ln \lambda_{\nu,k}^0 &= \sum_{m=1}^n \frac{\nu_{km}}{\nu_k} \ln \lambda_m^0 ,
\end{align}
puts the mean molar component activity coefficients in terms of species activity coefficients. Note that the versions of equations \eqref{meanmolaractivitydef} and \eqref{refchempotdef} written in terms of summations identify with the definitions stated by both Guggenheim \cite{Guggenheimbook} and Newman \cite{newman2004electrochemical}. 

\subsection{Electroneutral thermodynamic factors} 
Applying electroneutrality to equations \eqref{GibbsDuhem0} and \eqref{EulerGpartition} reveals equilibrium energetics to be independent of $\Phi_z^0$. The electroneutral Euler and Gibbs--Duhem equations, 
\begin{equation}\label{ElectroneutEulerG}
\tilde{G} = \left( \boldsymbol{\mu}_\nu^0 \right)^\top \textbf{c}_\nu 
\end{equation}
and
\begin{equation}\label{ElectroneutGD}
\textbf{c}_\nu^\top \nabla \boldsymbol{\mu}_\nu^0 = \vec{0} ,
\end{equation}
respectively, imply that isothermal, isobaric component chemical potentials have the functionality $\boldsymbol{\mu}_\nu^0 = \boldsymbol{\mu}_\nu^0 \left( \textbf{c}_\nu \right)$, or, equivalently, $\boldsymbol{\mu}_\nu^0 \left( \textbf{y}_\nu \right)$.\footnote{The implication $\boldsymbol{\mu}_\nu^0 \left( \textbf{c}_\nu \right) \implies \boldsymbol{\mu}_\nu^0 \left( \textbf{y}_\nu \right)$ relies on the fact that total concentration depends only on the species particle fractions at fixed temperature and pressure. This is justified by Euler equation \eqref{basicEulerG} because the molar Gibbs energy $\overline{G} = \tilde{G} / c_\textrm{T}$ must produce the molar volume $1/c_\textrm{T}$ through the derivative $(\partial \overline{G} / \partial p)_{T,y_i} = 1/c_\textrm{T}$.} The mean molar activity coefficients in electroneutral constitutive laws \eqref{munu0_constitutive} thus depend only on $\textbf{y}_\nu$ at constant temperature and pressure. 

These facts come together with the discussion of Darken factors from section \ref{sec:DarkenZ} to establish Maxwell relations and component thermodynamic factors for the electroneutral Onsager--Stefan--Maxwell diffusion driving forces. Let
\begin{equation}\label{Lambdanu0ijdef}
\Lambda_{\nu,ij}^0 =  \nu_i  \left( \frac{\partial \ln \lambda_{\nu,i}^0 }{\partial y_{\nu,j}} 
\right)_{T,p,y_{\nu,k \ne j,n-1}}
\end{equation}
represent the partial derivative of the mean molar activity coefficient of component $i$ with respect to the fraction of component $j$, leaving all of the $n-1$ component fractions save the $j$th and $(n-1)$th fixed. In light of Gibbs--Duhem equation \eqref{ElectroneutGD}, the component chemical-potential gradients can be expanded as
\begin{align}\label{gradmunu0ofLambda}
\nabla \boldsymbol{\mu}_\nu^0 = &  \nabla \boldsymbol{\mu}_\nu^{0,\textrm{ideal}} \nonumber \\
& + RT \left( \textbf{I}_{n-1} - \boldsymbol{\nu} \textbf{y}_\nu^\top \right)  \left[ \begin{array}{cc}
\boldsymbol{\Lambda}_\nu^0 & \textbf{o} \\ 
\textbf{o}^\top & 0
\end{array} \right] \nabla \textbf{y}_\nu,
\end{align}
in which $\boldsymbol{\Lambda}_\nu^0$ is an $(n-2) \times (n-2)$ matrix of electroneutral component Darken factors, and
\begin{align}\label{munu0idealdef}
\nabla \boldsymbol{\mu}_\nu^{0,\textrm{ideal}} &= RT \textbf{N}^\top \diag\left( \textbf{y}^0 \right)^{-1} \nabla \textbf{y}^0 \nonumber \\
&= RT \textbf{N}^\top \diag\left( \textbf{N} \textbf{y}_\nu \right)^{-1} \textbf{N} \nabla \textbf{y}_\nu
\end{align}
expresses the gradient of the ideal contribution to every component chemical potential.

Equation \eqref{gradmunu0ofLambda} contains $(n-2)^{2}$ excess component Darken factors $\Lambda^{0}_{\nu,ij}$, but these are not all independent because of  Maxwell relations, whose structure can be understood by revisiting the electroneutral form of equation \eqref{gradmunu}. After insertion of equation \eqref{munu0idealdef} and simplification with equation \eqref{y0ofynu}, equation \eqref{gradmunu} becomes
\begin{equation}\label{gradmuno0ofK}
\nabla \boldsymbol{\mu}_\nu^0 = \nabla \boldsymbol{\mu}_\nu^{0,\textrm{ideal}} + RT \left( \textbf{I}_{n-1} - \boldsymbol{\nu} \textbf{y}_\nu^\top \right) \Delta \textbf{K}_\nu^0 \nabla \textbf{y}_\nu,
\end{equation}
in which the $(n-1)$ by $(n-1)$ matrix $\Delta \textbf{K}_\nu^0$ is formed by striking the $n$th row and $n$th column of $\Delta \textbf{K}_Z^0$,
\begin{equation}
\Delta \textbf{K}_\nu^0 = \left[ \begin{array}{cc} 
\textbf{I}_{n-1} & \textbf{o} 
\end{array}
\right] \Delta \textbf{K}_Z^0 \left[ \begin{array}{c}
\textbf{I}_{n-1} \\
\textbf{o}^\top 
\end{array}\right],
\end{equation}
and is consequently symmetric. Equating the right sides of equations \eqref{gradmunu0ofLambda} and \eqref{gradmuno0ofK} shows that 
\begin{equation}\label{LambdaandK}
\left( \textbf{I}_{n-1} - \boldsymbol{\nu} \textbf{y}_\nu^\top \right)  \left[ \begin{array}{cc}
\boldsymbol{\Lambda}_\nu^0 & \textbf{o} \\ 
\textbf{o}^\top & 0
\end{array} \right]  = 
 \left( \textbf{I}_{n-1} - \boldsymbol{\nu} \textbf{y}_\nu^\top \right) \Delta \textbf{K}_\nu^0,
\end{equation}
which relates the thermodynamic factors that make up $\boldsymbol{\Lambda}_\nu^0$ to the truncated excess Hessian $\Delta \textbf{K}_\nu^0$. Compatibility of the species and component perspectives thus demands that the last column of $\Delta \textbf{K}_\nu^0$ lies in the nullspace of $\left( \textbf{I}_{n-1} - \boldsymbol{\nu} \textbf{y}_\nu^\top \right)$, that is, it must be proportional to $\boldsymbol{\nu}$.  (Guggenheim's principle demands that the constant of proportionality cannot be determined, however.) The last row of $\Delta \textbf{K}_\nu^0$ is therefore proportional to $\boldsymbol{\nu}^\top$, as required by the Hessian's symmetry. Multiplication through by $\textbf{y}_{\nu}^\top$  additionally verifies that the last row of equation \eqref{LambdaandK} is linearly dependent, since Gibbs--Duhem relations constrain both sides of the equality in the same way. 

In short, the last row of equation \eqref{LambdaandK} is redundant, and the last column, trivial. Discarding these, isolating the truncated Hessian that remains by exploiting a Sherman--Morrison formula, and further rearranging demonstrate that for all $i \ne j$, the Maxwell relations
\begin{equation}
\nu_i \left( \frac{\partial \ln \lambda_{\nu,i}^0 }{\partial y_{\nu,j} }\right)_{T,p,y_{\nu,k \ne j,n-1}} = \nu_j \left( \frac{\partial \ln \lambda_{\nu,j}^0 }{\partial y_{\nu,i} }\right)_{T,p,y_{\nu,k \ne i,n-1}}
\end{equation}
hold; the matrix $\boldsymbol{\Lambda}^{0}_{\nu}$ is symmetric. This substantiates the claim that $\tfrac{1}{2} \left( n-1 \right) \left( n-2 \right)$ properties, which quantify isothermal, isobaric gradients of $n-1$ component chemical potentials, parametrize the mixing free energy of any $n$-ary electroneutral electrolyte.

Henceforth, we will write the electroneutral gradients of component chemical potentials in the more compact form
\begin{equation}\label{di_of_ynu}
\nabla \boldsymbol{\mu}_\nu^0 = RT \textbf{X}_\nu^0 \nabla \textbf{y}_\nu ,
\end{equation}
in which the $(n-1) \times (n-1)$ matrix of \emph{electroneutral thermodynamic factors} $\textbf{X}_\nu^0$ is defined as
\begin{align}\label{Xnu0def}
\textbf{X}_\nu^0 = &\textbf{N}^\top \diag \left( \textbf{N} \textbf{y}_\nu \right)^{-1} \textbf{N} \nonumber \\ 
& +\left( \textbf{I}_{n-1} - \boldsymbol{\nu} \textbf{y}_\nu^\top \right)  \left[ \begin{array}{cc} 
\boldsymbol{\Lambda}_\nu^0 & \textbf{o} \\ 
\textbf{o}^\top & 0
\end{array} \right],
\end{align}
wherein $\boldsymbol{\Lambda}^{0}_{\nu} = \left( \boldsymbol{\Lambda}^{0}_{\nu}\right)^\top$, and equation \eqref{Lambdanu0ijdef} establishes how independent entries in $\boldsymbol{\Lambda}_\nu^0$ can be measured experimentally.

\subsection{Electroneutrality and dynamics} \label{subsec:Elec_and_Dyn}
Application of the electroneutrality approximation to the dynamical governing equations is simpler than electroneutral thermodynamics. When excess charge density vanishes, there is no change to the $(n-1)$-dimensional component material balance equation \eqref{cnubalances} that was produced by moving to a salt--charge basis. Charge continuity, embodied by equation \eqref{chgcont}, does simplify slightly, to
\begin{equation}\label{Kirchhoff_node}
\nabla \cdot \vec{i} = 0, ~~~~~~~~\textrm{(electroneutral)}
\end{equation}
which expresses Kirchhoff's law of the node. 

Entries in the Onsager drag matrix from equation \eqref{transport matrix} change their structure under electroneutrality. The transport matrix $\textbf{M}^0$ that relates species fluxes to species driving forces in the electroneutral case has entries
\begin{equation} \label{neutral_transport_matrix}
M_{ij}^0 = \begin{cases} -\dfrac{RT }{c_{\text{T}}^0 \mathscr{D}_{ij} }  \:\:\: \text{if} \:\: i \neq j \vspace{6pt} \\
\dfrac{RT}{c_{\text{T}}^0 } \displaystyle{ \sum_{k \neq i}^{n}} \dfrac{c_{k}^0}{\mathscr{D}_{ik} c_{j}^0} \:\:\: \text{if} \:\: i = j, \end{cases}
\end{equation}
where the neutral composition descriptors $\textbf{c}^0$ and $c_\textrm{T}^0$ are given in terms of the component concentrations $\textbf{c}_\nu$ by equations \eqref{c0def} and \eqref{cT0def}, respectively. Although the $n$ species concentrations here depend only on $n-1$ component concentrations from a salt--charge basis, the effect is relatively minor because the constraint does not affect the rank of $\textbf{M}^0$. 

When the whole set of Onsager--Stefan--Maxwell equations is put over a salt--charge basis, the transport coefficients that appear after applying electroneutrality follow from congruence relation \eqref{MZofM}:
\begin{equation}\label{MZ0block}
\textbf{M}_Z^0 = \textbf{Z} \textbf{M}^0 \textbf{Z}^\top = \left[ \begin{array}{cc} 
\textbf{M}_\nu & \textbf{m}_{z} \\
\textbf{m}_{z}^\top & M_{zz}
\end{array} \right].
\end{equation}
\noindent The second equality here defines sub-blocks that will be useful for expressing other macroscopic transport properties in terms of the electroneutral Onsager drag coefficients over a salt--charge basis, $\textbf{M}_Z^0$. Sub-block $\textbf{M}_\nu$ is an $\left(n-1\right) \times \left(n-1\right)$ square matrix, $\textbf{m}_{z}$, an $(n-1)$-dimensional column matrix, and $M_{zz}$, a scalar.

Recall from section \ref{sec:GEs_saltcharge} that the change to a salt--charge basis implies $\textbf{c}_Z$ is the sole null eigenvector of $\textbf{M}_Z$. After electroneutrality is adopted, the column $\textbf{c}_Z^0$, defined as
\begin{equation}
\textbf{c}_Z^0 = \left[ \begin{array}{c}
\textbf{c}_\nu \\
0
\end{array}\right],
\end{equation}
is a null eigenvector of $\textbf{M}_Z^0$. This in turn implies that
\begin{equation}
\textbf{M}_\nu \textbf{c}_\nu = \textbf{o},
\end{equation}
which is to say, the column of component concentrations is a null eigenvector of the symmetric sub-block 
$\textbf{M}_\nu$, and also that
\begin{equation}
\textbf{m}_{z}^\top \textbf{c}_\nu = 0,
\end{equation}
so that the column $\textbf{m}_{z}$ is orthogonal to the component concentrations.

Once the transport matrix from equation \eqref{MZ0block} has been incorporated, the Onsager--Stefan--Maxwell equations over a salt--charge basis from equation \eqref{OSM_saltcharge} become
\begin{equation}\label{OSM_neutral}
- \left[ \begin{array}{c}
\nabla \boldsymbol{\mu}_\nu^0 \\
F \left\| \textbf{z} \right\| \nabla \Phi_z^0 
\end{array}
\right] = \left[ \begin{array}{cc} 
\textbf{M}_\nu & \textbf{m}_{z} \\
\textbf{m}_{z}^\top & M_{zz}
\end{array} \right] \left[ \begin{array}{c} 
\vec{\textbf{n}}_\nu \\
\dfrac{\vec{i}}{F\left\| \textbf{z} \right\|}
\end{array} \right]
\end{equation}
under local electroneutrality. The transport matrix here is symmetric positive semidefinite and affords the single null eigenvector $\textbf{c}_Z^0$, preserving the spectral structure of the species flux laws. Bear in mind that the component chemical potential gradients $\nabla \boldsymbol{\mu}_\nu^0$ are parametrized by a set of thermodynamic factors $\textbf{X}_\nu^0$ dependent on $\textbf{y}_\nu$ and a matrix of material parameters through equations \eqref{Xnu0def}, in turn constrained by the symmetry of the excess property matrix $\boldsymbol{\Lambda}_\nu^0$. Henceforth we call equation \eqref{OSM_neutral}, with $\nabla \boldsymbol{\mu}_\nu^0$ parametrized by equations \eqref{di_of_ynu} and \eqref{Xnu0def}, the \emph{electroneutral Onsager--Stefan--Maxwell equations}. 

\section{Mass continuity and excess fluxes} \label{sec:convection}
Compatibility of the material balances with the familiar mass continuity equation is ensured by identifying the mass-average (barycentric) velocity as a combination of the species fluxes. Developing excess fluxes relative to the mass-average velocity allows an understanding of the non-dissipative phenomenon of convection, which is distinct from the dissipative diffusion process. 

To incorporate mass into the governing framework, let $\overline{m}_i$ represent the molar mass of species $i$, and assemble the set of these masses into an $n$-dimensional column $\overline{\textbf{m}}$. The mass density $\rho$ of a multi-species solution is generally given by
\begin{equation}\label{rhodef}
\rho = \sum_{i=1}^n \overline{m}_i c_i = \overline{\textbf{m}}^\top \textbf{c}.
\end{equation}
\noindent Further let $\psi_i$ represent a weighting factor with units of inverse molarity (or, equivalently, molar volume), 
\begin{equation}\label{psidef}
\psi_i = \frac{\overline{m}_i}{\rho},~~~~~\textrm{or}~~~~~\boldsymbol{\psi} = \frac{\overline{\textbf{m}}}{\rho},
\end{equation}
where the second relationship summarizes the definition for all $i$ by introducing an $n$-dimensional column matrix $\boldsymbol{\psi}$ with entries $\psi_i$. Note that equation \eqref{rhodef} implies a general constraint among the weighting factors, 
\begin{equation}\label{psiconstraint}
\boldsymbol{\psi}^\top \textbf{c} = 1,
\end{equation}
that is, the species mass fractions $\psi_i c_i$ sum to unity. The mass-average velocity $\vec{v}$ is generally defined as
\begin{equation}\label{vdef}
\vec{v} = \sum_{i=1}^n \psi_i \vec{N}_i = \boldsymbol{\psi}^\top \vec{\textbf{n}}.
\end{equation}
\noindent With equations \eqref{rhodef}, \eqref{psiconstraint}, and \eqref{vdef} in hand, one can multiply equation \eqref{speciescontinuities} through from the left by the constant row matrix $\rho \boldsymbol{\psi}^\top$ ($= \overline{\textbf{m}}^\top$) to get
\begin{equation}
\frac{\partial \rho}{\partial t} = - \nabla \cdot \left( \rho \vec{v} \right) ,
\end{equation}
showing consistency of the material balances with the common understanding of mass continuity. 

Defining a bulk velocity such as $\vec{v}$ distinguishes convective mass transport from diffusion as follows. Let the excess molar flux of species $i$ relative to the mass-average velocity, $\vec{J}_i$, be defined as
\begin{equation}\label{jdef}
\vec{J}_i = \vec{N}_i - c_i \vec{v},~~~~~\textrm{or}~~~~~\vec{\textbf{j}} = \vec{\textbf{n}} -  \vec{v} \, \textbf{c},
\end{equation}
where the latter equation introduces the $n$-dimensional column $\vec{\textbf{j}}$ whose entries are each of the $\vec{J}_i$ in sequence. Insertion of definition \eqref{jdef} into equation \eqref{OSMmatrix} gives
\begin{equation}\label{OSMofJ}
- \nabla \boldsymbol{\mu} = \textbf{M} \left( \vec{\textbf{j}} + \vec{v} \, \textbf{c} \right) = \textbf{M} \vec{\textbf{j}},
\end{equation}
which is to say, because $\textbf{c}$ is a null eigenvector of $\textbf{M}$, the physical content of the Onsager--Stefan--Maxwell equations is independent of convection. 

Importantly, the very presence of a bulk velocity constrains excess fluxes. In light of equations \eqref{psiconstraint} and \eqref{vdef}, multiplying equation \eqref{jdef} through by $\boldsymbol{\psi}^\top$ reveals that
\begin{equation}\label{kinematic}
\boldsymbol{\psi}^\top \vec{\textbf{j}} = \vec{0},
\end{equation}
establishing a kinematic relation among the set of excess molar species fluxes relative to the barycentric velocity.

Electroneutrality affects density and the kinematic relation in different ways. Application of electroneutrality to equation \eqref{rhodef} and insertion of equation \eqref{c0def} show that density is a function of $n-1$ component concentrations, 
\begin{equation}\label{rho0def}
\rho^0 = \overline{\textbf{m}}^\top \textbf{c}^0 = \overline{\textbf{m}}_\nu^\top \textbf{c}_\nu,
\end{equation}
in which $\overline{\textbf{m}}_\nu = \textbf{N}^\top \overline{\textbf{m}}$ is an $(n-1)$-dimensional column comprising the component molar masses. Excess species fluxes then transform into the salt--charge basis as
\begin{equation}\label{jZdef}
\vec{\textbf{j}}_Z = \textbf{Z}^{-\top} \vec{\textbf{j}} = \left[ \begin{array}{c} 
\vec{\textbf{j}}_\nu \\
\dfrac{\vec{i} - \rho_\textrm{e} \vec{v} }{F \left\| \textbf{z} \right\| }
\end{array} \right] ,
\end{equation}
because $\rho_\textrm{e} = F \textbf{z}^\top \textbf{c}$. Imposing local electroneutrality on $\vec{\textbf{j}}_Z$ here shows that the current density, embodied by the last entry of $\vec{\textbf{j}}_Z^{\,0}$, becomes independent of convection. This entry does not vanish entirely, however, and consequently electroneutrality does not reduce the dimensionality of the space of excess fluxes in the same way that it reduces the domain of $\rho^0$. 

Defining $\boldsymbol{\psi}_{Z} = \textbf{Z} \boldsymbol{\psi}$, one can change basis to write kinematic relation \eqref{kinematic} as
\begin{equation}
\boldsymbol{\psi}_{Z}^{\top} \vec{\textbf{j}}_Z = \vec{0},
\end{equation}
showing that the column of excess component fluxes is orthogonal to $\boldsymbol{\psi}_{Z}^{\top}$. Electroneutral species fluxes $\vec{\textbf{j}}^{\,0}$ depend on the component excess fluxes over a salt--charge basis, $\vec{\textbf{j}}_\nu$, as
\begin{equation}\label{j0ofjnu}
\vec{\textbf{j}}^{\,0} = \textbf{Z}^\top \vec{\textbf{j}}_Z^{\,0} = \textbf{N} \vec{\textbf{j}}_\nu + \dfrac{\vec{i} }{F \left\| \textbf{z} \right\| } \cdot \dfrac{\textbf{z}}{\left\| \textbf{z} \right\|} ,
\end{equation}
a form that notably includes the current, as well as the excess component fluxes. 

\section{Flux-explicit transport laws} 
Typical constitutive laws for electrolytic transport, including the standard concentrated solution theory and the Nernst--Planck equations, are written in flux-explicit forms. Laws giving component flux and current density in terms of the gradients of composition and voltage can be derived from force-explicit Onsager--Stefan--Maxwell equations by exploiting the inversion formulas reported previously by the authors  \cite{VanbruntAIChE}. Generally, flux-explicit laws for diffusion must be expressed using excess species fluxes in order to exclude the effects of convection; transport-law inversion therefore necessitates the selection of a representative bulk velocity. 

Section \ref{sec:convection} set out the preliminary definitions needed to develop flux-explicit transport laws relative to the mass-average velocity by the inversion method of Van-Brunt et al.\ \cite{VanbruntAIChE}. This method begins with a statement of the electroneutral Onsager--Stefan--Maxwell equations \eqref{OSM_neutral} in terms of excess fluxes. Through equation \eqref{jZdef}, they become
\begin{equation}\label{neutOSMofJ}
- \left[ \begin{array}{c}
\nabla \boldsymbol{\mu}_\nu^0 \\
F \left\| \textbf{z} \right\| \nabla \Phi_z^0 
\end{array}
\right] = \left[ \begin{array}{cc} 
\textbf{M}_\nu & \textbf{m}_{z} \\
\textbf{m}_{z}^\top & M_{zz}
\end{array} \right] \left[ \begin{array}{c} 
\vec{\textbf{j}}_\nu \\
\dfrac{\vec{i}}{F\left\| \textbf{z} \right\|}
\end{array} \right]
\end{equation}
in terms of $\vec{\textbf{j}}_\nu$. To implement the inversion, we first observe that within the flux-explicit form of these laws, the electroneutral excess component fluxes $\textbf{j}_\nu$ and $\vec{i}$ must be constrained such that ${\boldsymbol{\psi}_Z^0}^\top \textbf{j}_Z^0 = 0$, wherein the kinematic relation over the salt--charge basis, $\boldsymbol{\psi}_Z^0$, is defined such that $\boldsymbol{\psi}_Z^0 = \textbf{Z} \boldsymbol{\psi}^0$. (Here $\boldsymbol{\psi}^0$ expresses the basic kinematic relation that constrains excess species fluxes relative to the barycentric velocity under local electroneutrality, identified by replacing $\rho$ with $\rho^0$ in equation \eqref{psidef}.) Therefore, the flux-explicit transport laws must involve a coefficient matrix $\textbf{L}_Z^0$ with null eigenvector $\boldsymbol{\psi}_Z^0$. This matrix is formed uniquely from $\textbf{M}_Z^0$ through the limit process
\begin{equation}\label{LZfromMZlimit}
\textbf{L}_Z^0 = \lim_{\gamma \to 0} \left( \textbf{M}_Z^0 +\frac{\boldsymbol{\psi}_{Z}^0 {\boldsymbol{\psi}_{Z}^0}^{\top}}{\gamma} \right)^{-1} ,
\end{equation}
or, equivalently, from the algebraic equation
\begin{equation}\label{LZfromMZalgebra}
\textbf{L}_Z^0 = \left( \textbf{M}_Z^0 + \gamma \boldsymbol{\psi}_{Z}^0 {\boldsymbol{\psi}_{Z}^0}^{\top}  \right)^{-1} -\frac{\textbf{c}_Z^0 \left( \textbf{c}_Z^0\right)^\top}{\gamma} ,
\end{equation}
whose result is identical for any nonzero value of the augmentation parameter $\gamma$  \cite{VanbruntAIChE}. After partitioning $\textbf{L}_Z^0$ into the block form
\begin{equation}\label{fluxexplicitOnsagerneut}
\textbf{L}_Z^0 = \left[ \begin{array}{cc} 
\textbf{L}_\nu & \textbf{l}_{z} \\
\textbf{l}_{z}^\top & L_{zz}
\end{array} \right],
\end{equation}
one finds
\begin{equation}\label{neutJofmu}
\left[ \begin{array}{c} 
\vec{\textbf{j}}_\nu \\
\dfrac{\vec{i}}{F\left\| \textbf{z} \right\|}
\end{array} \right] 
=
- \left[ \begin{array}{cc} 
\textbf{L}_\nu & \textbf{l}_{z} \\
\textbf{l}_{z}^\top & L_{zz}
\end{array} \right] 
 \left[ \begin{array}{c}
\nabla \boldsymbol{\mu}_\nu^0 \\
F \left\| \textbf{z} \right\| \nabla \Phi_z^0 
\end{array}
\right] ,
\end{equation}
the desired flux-explicit form of the electroneutral Onsager transport laws with respect to a salt--charge basis.

\section{Distinguishing ionic conductivity, migration, and molecular diffusion}

One can leverage the flux-explicit electroneutral Onsager transport laws to generalize the process by which concentrated solution theory distinguishes the various dissipative phenomena that drive electrochemical mass transport \cite{newman2004electrochemical}. First and foremost, the last row of equation \eqref{neutJofmu} establishes the MacInnes equation---a current--voltage relation that can be viewed as a modified form of Ohm's law accounting for concentration overpotential \cite{macinnes1961principles}. Identifying the \emph{ionic conductivity} $\kappa^0$ as
\begin{equation}\label{conducdef}
\kappa^0 = F^2 \left\| \textbf{z} \right\|^2 L_{zz},
\end{equation}	
and further letting an $(n-1)$-entry column $\boldsymbol{\xi}$ represent a set of what we will call \emph{component migration coefficients}, defined by
\begin{equation}\label{xidef}
\boldsymbol{\xi} = \frac{1}{L_{zz}}  \textbf{l}_z,
\end{equation}
the MacInnes equation can be written as
\begin{equation}\label{MacInnes}
\vec{i} = - \kappa^0 \nabla \Phi_z^0 -   \frac{\kappa^0 \boldsymbol{\xi}^\top  \nabla \boldsymbol{\mu}_\nu^0}{F\left\| \textbf{z} \right\| } 
\end{equation}
in terms of the salt--charge potential and the component chemical potentials with respect to the chosen salt--charge basis. The first term on the right describes charge flow by ionic conduction, i.e., current driven by electric-potential gradients, in the absence of composition gradients.

Migration is defined as the excess mass flux driven by electric current flow in the absence of composition gradients. To isolate this effect, one writes the species flux laws in a partially inverted form, with current density replacing the electric-potential gradient. Newman does this by eliminating one of the species fluxes with Faraday's law \eqref{FaradayForCurrent} before performing a pseudoinversion of the electroneutral Onsager--Stefan--Maxwell equations \cite{newman2004electrochemical}. Equivalently, one can rearrange the Onsager flux laws from equation \eqref{neutJofmu}, by using the MacInnes equation---the last row of equation \eqref{neutJofmu}---to replace $\nabla \Phi_z^0$ with $\vec{i} / F \left\| \textbf{z} \right\|$ in the laws for $\vec{\textbf{j}}_\nu$ expressed by the remaining rows. Let the symmetric matrix
\begin{equation}
\overline{\textbf{L}}_\nu = \frac{RT}{c_\textrm{T}^0} \left( \textbf{L}_\nu - \frac{\kappa^0 \boldsymbol{\xi} \boldsymbol{\xi}^\top}{F^2 \left\| \textbf{z} \right\|^2} \right)
\end{equation}
define the set of \emph{Onsager component diffusivities} relative to the mass-average velocity, $\overline{\textbf{L}}_\nu$. The excess component fluxes become
\begin{equation}\label{componentFick}
\vec{\textbf{j}}_\nu = -   \frac{c_\textrm{T}^0 \overline{\textbf{L}}_\nu}{RT} \nabla \boldsymbol{\mu}_\nu^0 + \boldsymbol{\xi} \frac{\vec{i}}{F \left\| \textbf{z} \right\| },
\end{equation}
in which $\overline{\textbf{L}}_\nu$ has the standard units of area per time. Here, the first term on the right describes the component flux associated with diffusion, and the second, migration.

Note that in the absence of current density, equation \eqref{componentFick} reduces to a generalized form of Fick's law that expresses component excess fluxes solely in terms of the symmetric component-diffusivity matrix $\overline{\textbf{L}}_\nu$. Miller deployed transport laws with this structure to describe solutions containing multiple dissolved salts in some of his experimental tests of the Onsager reciprocal relations \cite{Miller1959}, which confirmed the symmetry of $\overline{\textbf{L}}_\nu$.

When quantifying migration coefficients, one should bear in mind that their values must be constrained to ensure that $\boldsymbol{\psi}_Z^0$ is a null eigenvector of $\textbf{L}_Z^0$. By partitioning $\boldsymbol{\psi}_{Z}^0$ into two parts $\boldsymbol{\psi}_\nu$ and $\psi_z$ such that
\begin{equation}\label{psipartition}
\boldsymbol{\psi}_{Z}^0 = \left[ \begin{array}{c}
\textbf{N}^\top \boldsymbol{\psi}^0 \\ 
\dfrac{\textbf{z}^\top \boldsymbol{\psi}^0}{\left\| \textbf{z} \right\| }
\end{array} \right] = \left[ \begin{array}{c}
\boldsymbol{\psi}_\nu \\ 
\psi_z
\end{array} \right],
\end{equation}
one can write
\begin{equation}\label{xiconstraint}
\boldsymbol{\psi}_\nu^\top \boldsymbol{\xi} = -\psi_z
\end{equation}
to phrase this constraint entirely in terms of species charges and molar masses. 

Within concentrated solution theory, transport laws are generally written for species fluxes, rather than component fluxes \cite{newman2004electrochemical}. By leveraging equation \eqref{j0ofjnu}, the species fluxes can be separated into terms associated with diffusion and migration using the properties defined above. Inserting equations \eqref{componentFick} yields
\begin{equation}\label{j0ofynu_and_i}
\vec{\textbf{j}}^{\,0} = -\frac{\textbf{N} \overline{\textbf{L}}_\nu c_\textrm{T}^0 }{RT} \nabla \boldsymbol{\mu}_\nu^0 +\left( \textbf{N} \boldsymbol{\xi}  + \frac{\textbf{z}}{\left\| \textbf{z} \right\|} \right) \frac{\vec{i}}{F \left\| \textbf{z} \right\| }.
\end{equation}
\noindent As well as distinguishing diffusion from migration at the species level, this expression helps to elucidate the spectral structure of the diffusivity matrix $\overline{\textbf{L}}_\nu$. Multiplying through from the left by $(\boldsymbol{\psi}^0)^\top$, applying the electroneutral form of equation \eqref{kinematic}, introducing $\boldsymbol{\psi}_\nu$ and $\psi_z$ with equation \eqref{psipartition}, and applying equation \eqref{xiconstraint} show that
\begin{equation}\label{Leigenvec}
-\frac{c_\textrm{T}^0 \boldsymbol{\psi}_\nu^\top \overline{\textbf{L}}_\nu}{RT} \nabla \boldsymbol{\mu}_\nu^0 = \vec{0}
\end{equation}
for any $\nabla \boldsymbol{\mu}_\nu^0$. Thus, because $\overline{\textbf{L}}_\nu$ is symmetric, it must afford $\boldsymbol{\psi}_\nu$ as a null eigenvector: $\overline{\textbf{L}}_\nu \boldsymbol{\psi}_\nu = \textbf{o}$.\footnote{If $\nabla \boldsymbol{\mu}_\nu^0$ is restricted to a space that satisfies the Gibbs--Duhem relation from equation \eqref{ElectroneutGD}, equation \eqref{Leigenvec} permits that $\overline{\textbf{L}}_\nu \boldsymbol{\psi}_\nu = k \boldsymbol{c}_\nu$ for some $k$. We choose $k = 0$ for consistency with the model formed by treating component transport in the absence of current as transport within a non-electrolytic system containing $n-1$ uncharged species, whose flux-explicit form matches equation \eqref{componentFick} with $\boldsymbol{\xi}=\textbf{o}$.}

A property count is helpful here. Symmetry of the $(n-1) \times (n-1)$ matrix $\overline{\textbf{L}}_\nu$ reduces its number of independent entries by $\tfrac{1}{2}\left(n-1\right) \left(n-2\right)$; the fact that it affords $\boldsymbol{\psi}_\nu$ as a null eigenvector adds $(n-1)$ additional constraints. Thus $\tfrac{1}{2}\left(n-1\right) \left(n-2\right)$ of the entries in $\overline{\textbf{L}}_\nu$ are independently specifiable. The $(n-1)$-dimensional column $\boldsymbol{\xi}$ is constrained through species molar masses and equivalent charges by equation \eqref{xiconstraint}, leaving $(n-2)$ independent migration coefficients. Finally, equation \eqref{conducdef} defines a single ionic conductivity. Summing up, the symmetric, positive-semidefinite flux-explicit Onsager matrix with respect to the salt--charge basis breaks down as
\begin{equation}\label{LZofDnuandxiandkappa}
\textbf{L}_Z^0 = \frac{c_\textrm{T}^0}{RT} \left[ \begin{array}{cc} 
\overline{\textbf{L}}_\nu & \textbf{o} \\
\textbf{o}^\top & 0 
\end{array} \right] + \frac{\kappa^0}{F^2 \left\| \textbf{z} \right\|^2} \left[ \begin{array}{cc} 
\boldsymbol{\xi} \boldsymbol{\xi}^\top & \boldsymbol{\xi} \\
\boldsymbol{\xi}^\top & 1 
\end{array} \right],
\end{equation}
in which $\overline{\textbf{L}}_\nu = \overline{\textbf{L}}_\nu^\top$, $\overline{\textbf{L}}_\nu \boldsymbol{\psi}_\nu = \textbf{o}$, and $\boldsymbol{\psi}_\nu^\top \boldsymbol{\xi} + \psi_z = 0$, with $\boldsymbol{\psi}_\nu$ and $\psi_z$ defined in terms of stoichiometry in the simple association equilibria, species molar masses, species charges, and the electroneutral density $\rho^0$ through equations \eqref{psidef} and \eqref{psipartition}. Consequently, $\textbf{L}_Z^0$ depends on $\tfrac{1}{2}n (n-1)$ independent transport coefficients. This equals the number of Stefan--Maxwell diffusivities that underpin the original transport matrix $\textbf{M}^0$ in equation \eqref{neutral_transport_matrix}. 

Recently molecular-dynamics techniques have been deployed to measure the naive Onsager diffusion matrix $\textbf{L}^0$, which sits relative to the mass-average velocity and appears in the inverted form of equation \eqref{OSMofJ} \cite{Fong2020, Mandadapu2020},\footnote{A molecular dynamics simulation can be described with equation \eqref{Lform} if the simulated control volume conserves charge and has characteristic dimensions much larger than the Debye length.}
\begin{equation}\label{Lform}
\vec{\textbf{j}}^{\,0} = - \textbf{L}^0 \nabla \boldsymbol{\mu}^0.
\end{equation}
\noindent The transport properties that make up $\textbf{L}^0$ relate to $\textbf{M}^0$ directly through
\begin{equation}
\textbf{M}^0 = \lim_{\gamma \to 0} \left[ \textbf{L}^0 + \frac{\textbf{c}^0 \left( \textbf{c}^0 \right)^\top}{\gamma} \right]^{-1},
\end{equation}
where we have used the limit form of the inversion described by Van-Brunt et al.\ \cite{VanbruntAIChE}. Through equation \eqref{LZofDnuandxiandkappa}, the congruence transformation
\begin{equation}
\textbf{L}_Z^0 = \textbf{Z}^{-\top} \textbf{L}^0 \textbf{Z}^{-1}
\end{equation}
shows how the component diffusivities, migration coefficients, and ionic conductivity over a salt--charge basis contribute to the naive Onsager matrix $\textbf{L}^0$.

\section{Alternative convective velocities}
Up to now we have adopted a convention of defining $\textbf{L}_{Z}^{0}$ and $\vec{\textbf{j}}^{\,0}$ relative to the mass-average velocity, but it is straightforward to change the kinematic constraint that defines the convective reference velocity in equation \eqref{vdef} to any other $\boldsymbol{\psi}'$ satisfying equation \eqref{psiconstraint}. (For example, letting $\boldsymbol{\psi}' = \textbf{1}/c_\textrm{T}$ defines the mole-average velocity.) The conversion of the set of excess molar fluxes relative to the mass-average velocity, composed of $\vec{J}_{i}$, to excess fluxes relative to another velocity, composed of $\vec{J}_{i}^{\,\psi'}$, is accomplished by the projection operation
\begin{equation}\label{jpsi_to_jpsiprime}
\vec{\textbf{j}}^{\psi'} = \left( \textbf{I} - \textbf{c} \boldsymbol{\psi}'^{\top} \right) \vec{\textbf{j}}.
\end{equation}
\noindent Observe that a similar projection can be used to move from excess fluxes that satisfy kinematic relation $\boldsymbol{\psi}'$ back to fluxes in excess of convection at the mass-average reference velocity:
\begin{equation}\label{jpsiprime_to_jpsi}
\vec{\textbf{j}} = \left( \textbf{I} - \textbf{c} \boldsymbol{\psi}^{\top} \right) \vec{\textbf{j}}^{\psi'}.
\end{equation}
\noindent Transformation \eqref{jpsi_to_jpsiprime} can be expressed in terms of excess fluxes over a salt--charge basis as
\begin{equation}\label{jZpsiprime}
\vec{\textbf{j}}^{\,\psi'}_{Z} = \left[ \textbf{I} - \textbf{c}_{Z} (\boldsymbol{\psi}'_{Z})^{\top} \right] \vec{\textbf{j}}_{Z},
\end{equation}
where $\boldsymbol{\psi}'_Z=\textbf{Z} \boldsymbol{\psi}'$. 

Under electroneutrality, the kinematic relation $\boldsymbol{\psi}_Z$ is replaced by $\boldsymbol{\psi}_Z^0$. Given any other kinematic relation $\boldsymbol{\psi}'$ defined under the electroneutrality approximation, one can exploit equation \eqref{LZfromMZlimit} or \eqref{LZfromMZalgebra}, with $\boldsymbol{\psi}'_Z$ in place of $\boldsymbol{\psi}_Z^0$, to identify the flux-explicit electroneutral transport relations relative to $\boldsymbol{\psi}'$, as
\begin{equation} \label{L_Z change of reference}
\vec{\textbf{j}}^{\,0,\psi'}_{Z}  = -\textbf{L}^{0,\psi'}_{Z} \nabla \boldsymbol{\mu}^{0}_{Z},
\end{equation}
in which
\begin{equation}\label{Lpsi}
\textbf{L}^{0,\psi'}_{Z} = \left[ \textbf{I} - \textbf{c}^{0}_{Z} (\boldsymbol{\psi}'_{Z})^{\top}
 \right] \textbf{L}^{0}_{Z} \left[ \textbf{I} -  \boldsymbol{\psi}'_{Z} (\textbf{c}^{0
 }_{Z})^{\top} \right]
\end{equation}
establishes a congruence between $\textbf{L}^{0,\psi'}_{Z}$ and $\textbf{L}^{0}_{Z}$. 
One can also compute $\textbf{L}^{0,\psi'}_Z$ directly from $\textbf{M}_Z^0$, using
\begin{equation}\label{Lpsiprime_of_MZ0}
\textbf{L}^{0,\psi'}_Z = \lim_{\gamma \to 0} \left( \textbf{M}_Z^0 + \frac{\boldsymbol{\psi}'_Z {\boldsymbol{\psi}'_Z}^\top }{\gamma} \right)^{-1},
\end{equation}
which is analogous to equation \eqref{LZfromMZlimit}. In the new reference frame for the excess fluxes established by $\boldsymbol{\psi}'$, diffusivities and migration coefficients can be identified by partitioning $\textbf{L}^{0,\psi'}_{Z}$ as 
\begin{align}\label{LZ0psiofDpsixipsi}
\textbf{L}_Z^{0,\psi'} = & \frac{c_\textrm{T}^0}{RT} \left[ \begin{array}{cc} 
\overline{\textbf{L}}_\nu^{\psi'} & \textbf{o} \\
\textbf{o}^\top & 0 
\end{array} \right] \nonumber \\ & + \frac{\kappa^0}{F^2 \left\| \textbf{z} \right\|^2} \left[ \begin{array}{cc} 
\boldsymbol{\xi}^{\psi'} \big( \boldsymbol{\xi}^{\psi'} \big)^\top & \boldsymbol{\xi}^{\psi'} \\
\big( \boldsymbol{\xi}^{\psi'}\big)^\top & 1 
\end{array} \right],
\end{align}
similar to how $\textbf{L}_Z^0$ is partitioned in equation \eqref{LZofDnuandxiandkappa}. Note that the second matrix on the right of equation \eqref{LZ0psiofDpsixipsi} is proportional to the outer product of $[ \boldsymbol{\xi}^{\psi'}, 1 ]^\top$ with itself, and is consequently positive semidefinite in general. 

 One can write excess component fluxes relative to the alternative convective velocity defined by $\boldsymbol{\psi}'$, and with respect to the salt--charge basis, as 
\begin{equation}\label{ComponentFick_psiprime}
\vec{\textbf{j}}_\nu^{\,\psi'} = -   \frac{c_\textrm{T}^0 \overline{\textbf{L}}_\nu^{\psi'}}{RT} \nabla \boldsymbol{\mu}_\nu^0 + \boldsymbol{\xi}^{\,\psi'} \frac{\vec{i}}{F \left\| \textbf{z} \right\| },
\end{equation}
which derives from equation \eqref{componentFick}.  Here, the component diffusivities relative to the average velocity determined by $\boldsymbol{\psi}'$ are defined as
\begin{equation} \label{General diffusivities}
\overline{\textbf{L}}_{\nu}^{\psi'} = \left[ \textbf{I}_{n-1} - \textbf{c}_{\nu} (\boldsymbol{\psi}'_{\nu})^{\top}
 \right] \overline{\textbf{L}}_{\nu} \left[ \textbf{I}_{n-1} -  \boldsymbol{\psi}'_{\nu} (\textbf{c}
_{\nu})^{\top} \right],
\end{equation}
and
\begin{equation}\label{xipsiprime}
\boldsymbol{\xi}^{\psi'} = \left[ \textbf{I}_{n-1} - \textbf{c}_{\nu} (\boldsymbol{\psi}'_{\nu})^{\top}
 \right] \boldsymbol{\xi} - \psi_{z}' \textbf{c}_{\nu} 
\end{equation}
defines migration coefficients in the new frame. Similar to the projection in equation \eqref{jpsiprime_to_jpsi}, one can create transformations that change the reference frame from velocity ${\boldsymbol{\psi}'}^\top \vec{\textbf{n}}$ back to the mass-average velocity by replacing $\boldsymbol{\psi}'_\nu$ with $\boldsymbol{\psi}_\nu$ and $\psi'_z$ with $\psi_z$ in equations \eqref{General diffusivities} and \eqref{xipsiprime}, as well as swapping $\overline{\textbf{L}}_\nu^{\psi'}$ with $\overline{\textbf{L}}_\nu$ and $\boldsymbol{\xi}^{\psi'}$ with $\boldsymbol{\xi}$. 

In Newman's implementations of concentrated solution theory \cite{newman2004electrochemical}, one of the species velocities---say, that of species $m$---is used as the convective reference. Thus the coefficients that Newman identifies in the flux-explicit equations follow from taking $\boldsymbol{\psi}'_Z = \textbf{Z} \textbf{i}_m/c^0_m$ in equation \eqref{Lpsi}. This produces species flux laws in the form
\begin{equation}\label{NewmanFluxLaws}
\vec{\textbf{j}}^{\,0,\psi'} = -\frac{c_\textrm{T}^0 \textbf{N} \overline{\textbf{L}}_\nu^{\psi'}}{RT} \nabla \boldsymbol{\mu}_\nu^0 +\left( \textbf{N} \boldsymbol{\xi}^{\psi'}  + \frac{\textbf{z}}{\left\| \textbf{z} \right\|} \right) \frac{\vec{i}}{F \left\| \textbf{z} \right\| }.
\end{equation}
\noindent Newman's development also produces a slightly different MacInnes equation,
\begin{equation}\label{NewmanMacInnes}
\vec{i} = - \kappa^0 \nabla \Phi_z^0 -  \frac{\kappa^0 }{F\left\| \textbf{z} \right\| }\big( \boldsymbol{\xi}^{\psi'} \big)^\top \nabla \boldsymbol{\mu}_\nu^0,
\end{equation}
wherein the migration coefficients are constrained by kinematic relation $\boldsymbol{\psi}'$, but $\kappa^0$ remains unaffected because its value is independent of the choice of convective velocity.

Note that the embedding of an arbitrary reference velocity in flux-explicit transport matrices (e.g., $\textbf{L}^0$) makes properties derived from them somewhat ambiguous. When tabulating transport parameters, it may be preferable to work with the electroneutral transport matrix $\textbf{M}^{0}_{Z}$, which conveys frame-invariant information. The computation of $\textbf{M}_{\nu}$ can be accomplished readily given a flux-explicit form; in fact, one can use the procedure of Van-Brunt et al.\ \cite{VanbruntAIChE} and a block inversion identity to show for any $\boldsymbol{\psi}'$ that
\begin{equation}\label{MnuofL}
\textbf{M}_{\nu} = \frac{RT}{c^{0}_{\textrm{T}}} \lim_{\gamma \rightarrow 0}  \left( \overline{\textbf{L}}^{\, \psi'}_{\nu} + \frac{ \textbf{c}_{\nu} \textbf{c}_{\nu}^{\top}}{\gamma} \right)^{-1},
\end{equation}
and also that 
\begin{equation} \label{Reconstruct M_Z}
\textbf{m}_{z} = \textbf{M}_{\nu} \boldsymbol{\xi}^{\psi'} ~~\textrm{and}~~ M_{zz} = \frac{F^2 \| \textbf{z} \|^2 }{\kappa^0} + (\boldsymbol{\xi}^{\psi'})^{\top} \textbf{M}_{\nu} \boldsymbol{\xi}^{\psi'},
\end{equation}
elucidating the structure of $\textbf{M}_{Z}^{0}$ through equation \eqref{MZ0block}. Despite its computation from a transport matrix $\textbf{L}_{Z}^{0,\psi'}$, it should be emphasized that $\textbf{M}_{Z}^{0}$ enjoys uniqueness and is independent of the convective velocity. Notably, the submatrix $\textbf{M}_{\nu}$ depends only on the Onsager component diffusivities, and the only matrix inversion that needs to be performed is the one in equation \eqref{MnuofL}.  The entries of $\textbf{m}_{z}$ are zero in the absence of migration, and $M_{zz}$ matches the ionic resistivity plus a quadratic correction that also vanishes when migration does. 

\section{Transference numbers}
Equations \eqref{MacInnes} and \eqref{j0ofynu_and_i} provide a pair of constitutive equations similar, but not identical, to the familiar equations from concentrated solution theory and Nernst--Planck dilute-solution theory. To achieve complete agreement, one must consider how migration coefficients relate to the transference numbers more familiar from electrochemistry \cite{macinnes1961principles, newman2004electrochemical, bard2000electrochemical}. Let the transference number of species $i$ relative to the mass-average velocity be $t_i$. These can be formed into an $n$-dimensional column matrix $\textbf{t}$, related to the migration coefficients as
\begin{equation} \label{tdef}
\textbf{t} = \frac{\diag\left(\textbf{z}\right)}{\left\| \textbf{z} \right\| } \left( \textbf{N} \boldsymbol{\xi} + \frac{\textbf{z}}{\left\| \textbf{z} \right\|} \right) .
\end{equation}
\noindent Multiplication of this result through by $\textbf{1}^\top$ shows that $\textbf{1}^\top \textbf{t} = 1$, that is, transference numbers sum to unity. Note that if migration is neglected entirely ($\boldsymbol{\xi}=\textbf{o}$), the right side of equation \eqref{tdef} remains non-zero; each transference number becomes a partition of unity weighted by the corresponding squared species charge. The migration coefficients we introduced here deviate from this standard.  

Sometimes it is desirable to change the reference velocity relative to which transference is expressed. Incorporating the alternative migration coefficient from equation \eqref{xipsiprime} into equation \eqref{tdef} shows that
\begin{equation}\label{tpsiprime}
\textbf{t}^{\psi'} = \frac{\diag\left(\textbf{z}\right)}{\left\| \textbf{z} \right\| } \left( \textbf{N} \boldsymbol{\xi}^{\psi'} + \frac{\textbf{z}}{\left\| \textbf{z} \right\|} \right)
\end{equation}
yields the set of species transference numbers relative to the reference velocity associated with the kinematic relation represented by $\boldsymbol{\psi}'$. These also afford the property that $\textbf{1}^\top \textbf{t}^{\psi'} = 1$.

Because $\diag\left( \textbf{z} \right)$ appears on the right of equation \eqref{tdef}, $t_i$ vanishes for all $i$ such that $z_i = 0$. Thus, transference numbers as traditionally defined only have meaning for charged species. Importantly, specifying a set of independent transference numbers does not generally suffice to specify the Onsager transport matrix. Newman addresses this point by using transference numbers for uncharged species expressed as ratios $t_i / z_i$, noting that the ratio is finite for any species $i$ with $z_i = 0$ \cite{newman2004electrochemical}. Indeed, one can see with equation \eqref{tdef} that $t_i/z_i$ can be written under the assumption that $t_i \propto z_i$ as 
\begin{equation} \label{Dividingbyzero}
\frac{t_i}{z_i} = \frac{1}{\left\| \textbf{z} \right\| } \left( \textbf{N} \boldsymbol{\xi} + \frac{\textbf{z}}{\left\| \textbf{z} \right\|} \right)_i,
\end{equation}
in which the right side is generally nonzero for any $i$. 

Rather than taxing ourselves by ignoring apparent divisions by zero when transference numbers are involved, we propose that it is simpler to think of transport fundamentally in terms of migration coefficients, because they are defined unambiguously in terms of Onsager coefficients for species with and without charge. The nontrivial transference numbers that make up $\textbf{t}$ are subsidiary quantities that afford a useful physical interpretation, but the set as a whole does not have as much essential physical content. The distinction between $\textbf{t}$ and $\boldsymbol{\xi}$ becomes particularly important for electrolytes containing multiple uncharged species, such as the cosolvent blends commonly used in lithium-ion-battery electrolytes \cite{Xu2004,Valoen2005}, fuel-cell membranes susceptible to gas crossover \cite{Weber2008}, or electrolytes for lithium--air batteries \cite{4componentLiO2}. Within these systems, applied currents can drive migration of one neutral species relative to another---a dissipative phenomenon distinct from diffusion that cannot be accounted for by traditional transference numbers.

\section{Component diffusion coefficients}
The entries of the Onsager component-diffusivity matrices $\overline{\textbf{L}}_\nu$ and $\overline{\textbf{L}}_\nu^{\,\psi'}$ have the expected units, and each entry is paired with a particular chemical-potential gradient. Nevertheless, the dependences of these transport matrices on the convective velocity obfuscates their connection to the energy dissipation. Newman addresses this issue by identifying combinations of the Stefan--Maxwell coefficients between species that quantify the frame-invariant part of the dissipation arising from component interdiffusion \cite{newman2004electrochemical,Newman-molten-salt1979}. The process Newman uses to identify thermodynamic component diffusivities within binary solutions can be generalized to solutions comprising more than two components as follows.

Instead of using entries of Onsager's diffusivity matrix $\overline{\textbf{L}}_\nu$, a frame-invariant set of thermodynamic component diffusivities is defined by revisiting the Stefan--Maxwell form. Physically, one expects that in situations where electroneutrality is maintained ($\rho_\textrm{e} = 0$) and the current density vanishes uniformly ($\vec{i}=\vec{0}$), the mass-transport model describing an $n$-species electrolytic phase should be no different than the model describing a corresponding $(n-1)$-species nonelectrolytic phase in which every uncharged species is taken to be a component of the original $n$-ary electrolyte. This equivalence suggests that $\textbf{M}_{\nu}$ possesses the structure 
\begin{equation}\label{component transport matrix}
\frac{c_{\text{T}}^{0}}{RT} \left( \textbf{M}_{\nu}\right)_{ij} = \begin{cases} - \dfrac{\nu_i \nu_j}{\mathfrak{D}_{ij}}  \:\:\: \text{if} \:\: i \neq j \vspace{6pt} \\
\displaystyle{ \sum_{k \neq i}^{n-1}} \dfrac{\nu_i \nu_k c_{\nu,k}}{\mathfrak{D}_{ik} c_{\nu,j}} \:\:\: \text{if} \:\: i = j \end{cases},
\end{equation}
which defines an alternative set of binary diffusion coefficients $\mathfrak{D}_{ij}$ that describe drag interactions between components $i$ and $j$ as they move relative to each other. It should be borne in mind that the indices in $\mathfrak{D}_{ij}$ refer to components in a given salt--charge basis.

In equation \eqref{component transport matrix}, the diffusivities that parametrize current-free, electroneutral component--component interactions are defined analogously to how species--species interactions are defined in equation \eqref{transport matrix}, with a minor stoichiometric correction to account for the fact that component $i$ within a salt--charge basis is actually made up of $\nu_i$ species through the simple association equilibria. We shall distinguish component diffusion coefficients from Stefan--Maxwell coefficients by writing them in Fraktur font, rather than a script font. 

Intrinsic properties of the Onsager matrix $\textbf{M}_Z$ demand that the coefficients defined by equation \eqref{component transport matrix} are symmetric in their indices, $\mathfrak{D}_{ij} = \mathfrak{D}_{ji}$, and that $\mathfrak{D}_{ii}$ is undefined for all $i$. Hence equation \eqref{component transport matrix} defines $\tfrac{1}{2} (n-1) (n-2)$ independent diffusion coefficients---one for each pair of components. The matrix $\textbf{M}_{\nu}$ must be positive semidefinite, a property guaranteed if each $\mathfrak{D}_{ij}$ is positive, and by construction it affords $\textbf{c}_\nu$ as a null eigenvector. Furthermore, a block inversion identity shows that equations \eqref{Lpsiprime_of_MZ0} and \eqref{LZ0psiofDpsixipsi} imply 
\begin{equation}\label{barL_of_barM}
\overline{\textbf{L}}_\nu^{\psi'} = 
\lim_{\gamma \to 0} \left( \frac{ c_{\text{T}}^{0}}{RT} \textbf{M}_{\nu} + \frac{\boldsymbol{\psi}'_\nu {\boldsymbol{\psi}'_\nu}^\top }{\gamma} \right)^{-1},
\end{equation}
an inverted form of equation \eqref{MnuofL}. Thus $\overline{\textbf{L}}_\nu^{\psi'}$, the portion of the electroneutral Onsager component-diffusivity matrix that describes situations in which current density vanishes uniformly, relates directly to $\textbf{M}_{\nu}$.

When casting $\overline{\textbf{L}}_\nu$ in terms of component diffusivities $\mathfrak{D}_{ij}$ with definition \eqref{component transport matrix} and inversion formula \eqref{barL_of_barM}, it is necessary to assume that every entry of $\textbf{c}_\nu$ is nonzero almost everywhere. This must be borne in mind when considering systems where products of the simple association equilibria are not unique.  The situation described in section \ref{sec:construction}, where two binary salts without a common ion are dissolved into an uncharged solvent, provides a simple example. That system can pass locally through compositions where a particular component concentration vanishes, but all of the ions retain non-negative concentrations, as described in the discussion at the end of section \ref{subsec:ElectroneutComp}. Such a state nominally induces singular behavior in one of the diagonal entries defined in equation \eqref{component transport matrix}. Observe, however, that the component diffusivities $\mathfrak{D}_{ij}$ are defined solely in terms of the off-diagonal entries of $\textbf{M}_{\nu}$. Therefore, although one must take care when applying inversion formula \eqref{barL_of_barM}, the diffusion coefficients $\mathfrak{D}_{ij}$ remain well posed across the physically available range of $\textbf{c}_\nu$. 

\section{Reference electrodes}
 
Introducing the concept of a reference electrode allows the salt--charge potential to be related unambiguously to an experimentally measurable voltage, despite the degree of freedom in $\Phi_z^0$ demanded by Guggenheim's principle. A heterogeneous half-reaction suitable for use within a reference electrode is generally written as
\begin{equation}\label{refreaction}
\sum_{i=1}^{n} s_{i} \textrm{X}_{i}^{z_i} + \sum_{j} s'_j \textrm{Y}_{j} \leftrightharpoons n_{\textrm{e}^-} \textrm{e}^{-},
\end{equation}
in which $\textrm{X}_i^{z_i}$ is the chemical symbol of species $i$, $\textrm{Y}_{j}$ is the symbol of uncharged entity $j$ that forms in an immediately adjacent phase, $\textrm{e}^-$ denotes electrons, the integer $s_i$ is the stoichiometric coefficient of species $i$ in the half-reaction, integer $s'_j$ is the stoichiometric coefficient of $\textrm{Y}_{j}$, and the positive integer $n_{\textrm{e}^-}$ is the number of electrons involved. Since half-reactions balance charge, the set $\{ s_{i} \}_{i=1}^{n}$ must generally satisfy
\begin{equation}\label{sz}
\sum_{i=1}^{n} s_{i} z_{i} = \textbf{s}^\top \textbf{z} = n_{\textrm{e}^-} z_{\textrm{e}^-},
\end{equation}
where $z_{\textrm{e}^-}$ is the equivalent charge of electrons. (If a half-reaction is written as a reduction, then $s_i$ or $s'_i$ is negative when species $i$ is a reactant and positive when it is a product.) Note that the matrix form in the middle of this expression indicates that by definition, the column $\textbf{s}$ representing half-reaction stoichiometry is not orthogonal to the column $\textbf{z}$ of species charges.

The electrode potential $\Phi^\ominus$ associated with reference-electrode reaction \eqref{refreaction} is defined such that \cite{newman2004electrochemical,Guggenheimbook}
\begin{equation}
n_{\textrm{e}^-} F z_{\textrm{e}^{-}} \nabla \Phi^\ominus = \sum_{i=1}^{n} s_{i} \nabla \mu_{i}.  
\end{equation}
\noindent After defining $\textbf{s}_{Z} = \textbf{Z}^{-\top} \textbf{s} = [\textbf{s}_{\nu}, s_{z} ]^{\top}$ and assuming an electroneutral state, this is written equivalently as
\begin{equation}
n_{\textrm{e}^-} F z_{\textrm{e}^{-}}  \nabla \Phi^\ominus  = \textbf{s}_{Z}^\top \nabla \boldsymbol{\mu}_{Z}^{0} = \textbf{s}_{\nu}^\top \nabla \boldsymbol{\mu}_\nu^0 +  s_z F \left\| \textbf{z} \right\| \nabla \Phi_z^0,
\end{equation}
and since
\begin{equation}
n_{\textrm{e}^-} z_{\textrm{e}^-} = \textbf{s}^\top \textbf{z} = \textbf{s}_Z^\top \textbf{Z}\textbf{z} = s_z \left\| \textbf{z} \right\|,
\end{equation}
it follows that 
\begin{equation}\label{nablaPhiominus}
 \nabla \Phi^\ominus  = \frac{ \textbf{s}_{\nu}^\top \nabla \boldsymbol{\mu}_\nu^0 }{n_{\textrm{e}^-} F z_{\textrm{e}^-}  } + \nabla \Phi_z^0,
\end{equation}
a relation that allows the voltage of a given reference electrode to be inferred after one has computed the component and salt-charge potentials. Note that at uniform temperature, pressure, and composition,
\begin{equation}
\left( \nabla \Phi^\ominus \right)_{T, p, \textbf{y}_\nu} = \left( \nabla \Phi_z^0 \right)_{T, p, \textbf{y}_\nu}
\end{equation}
for any reference-electrode half-reaction, showing the intrinsic utility of the salt--charge potential.

\section{Examples}

\subsection{The binary electrolyte}\label{subsec:binaryelectrolyte}
The model of a binary electrolytic solution---a liquid comprising a single simple salt dissolved in a single uncharged solvent---is the simplest and most common implementation of concentrated solution theory. Newman's identification of transport laws for this system begins by adopting Onsager--Stefan--Maxwell equations \eqref{OSMequation} with the transport coefficients defined in equation \eqref{transport matrix}. He uses these in combination with the electroneutrality approximation \eqref{electroneutrality.appx}, Guggenheim condition \eqref{Guggenheim.condition} for the salt, and Faraday's law \eqref{FaradayForCurrent} to derive species flux laws in the form of equation \eqref{NewmanFluxLaws} and a MacInnes equation in the form of equation \eqref{NewmanMacInnes}, casting both in terms of properties relative to the solvent velocity. The development involves apparently ad hoc manipulation of the transport equations to obtain compact forms of the flux-explicit laws.

We now demonstrate how Newman's process can be formalized with a salt--charge basis under electroneutrality. Let $0$, $+$, and $-$ distinguish the solvent, cation, and anion species, respectively, and subscripts $0, \text{e}$ denote the solvent and salt components. Set the stoichiometric coefficients for the reactants in the simple association equilibrium forming the salt as
\begin{equation}\label{binarystoich}
\nu_{+} = -\frac{\nu z_{-}}{z_{+} - z_{-}}, \:\:\:\:\:\: \nu_- = \frac{\nu z_{+}}{z_{+} - z_{-}},
\end{equation}
and note that $\nu = \nu_{+} + \nu_{-}$.
This framing demonstrates that $z_+ \nu_+ + z_- \nu_- = 0$ more directly, and will produce results more obviously consistent with Newman \cite{NEWMAN1997}. 
The products of the simple association equilibria for this system are unique, so the salt--charge basis is unique up to ordering of the ions. We let the salt--charge transformation $\textbf{Z}$ take the form
\begin{equation}\label{Z0+-}
\textbf{Z} =  \begin{bmatrix}
1 & 0 & 0 \\ 0 & -\frac{\nu z_{-}}{z_{+} - z_{-}} &  \frac{\nu z_{+}}{z_{+} - z_{-}} \\
0 & \frac{z_{+}}{\left\| \textbf{z} \right\|} & \frac{z_{-}}{\left\| \textbf{z} \right\|} 
\end{bmatrix},
\end{equation}
in which $\left\| \textbf{z} \right\| = \sqrt{z_+^2 + z_-^2}$ because $z_0 = 0$. Given this, one can write the set of species concentrations in the neutral state, $\textbf{c}^0$, as a parametric function of the component concentrations $c_0$ and $c_\textrm{e}$ through equation \eqref{c0def}, 
\begin{equation}
\left[ \begin{array}{c}
c_0^0 \\
c_+^0 \\
c_-^0 
\end{array} \right] = \textbf{Z}^\top \left[ \begin{array}{c}
c_0 \\
c_\textrm{e} \\
0
\end{array} \right] = \left[ \begin{array}{c}
c_0 \\
- \frac{z_- \nu }{z_+ - z_-} \cdot c_\textrm{e}  \\
\frac{z_+ \nu }{z_+ - z_-} \cdot c_\textrm{e}
\end{array}
\right].
\end{equation} 
\noindent The electroneutral component chemical potentials can be written using equations \eqref{di_of_ynu} and \eqref{Xnu0def}, which together with the relation $y_0 = 1- \nu y_\textrm{e}$ show that
\begin{align}
\nabla \left[ \begin{array}{c}
\mu_0^0 \\
\mu_\textrm{e}^0
\end{array} \right] &= RT\left[ \begin{array}{cc}
X_{00}^0 & X_{0\textrm{e}}^0 \\
X_{\textrm{e}0}^0 & X_{\textrm{ee}}^0
\end{array} \right] \nabla \left[ \begin{array}{c}
y_0 \\
y_\textrm{e}
\end{array} \right] \nonumber \\
&= \frac{RT \left( 1 + \Lambda_{00} \nu y_\textrm{e} y_0 \right)}{y_\textrm{e} y_0 } \left[ \begin{array}{c}
-y_\textrm{e} \\
y_0 
\end{array} \right] \nu \nabla y_\textrm{e},
\end{align}
parametrized by the single excess activity-coefficient derivative function $\Lambda_{00} \left( y_\textrm{e} \right)$.

Using the salt--charge transformation from equation \eqref{Z0+-} in equation \eqref{MZ0block} shows that
\begin{equation}\label{Mnu0+-}
\textbf{M}_\nu = \frac{\nu RT \left( z_+ \mathscr{D}_{0+} - z_- \mathscr{D}_{0-} \right) }{c_\textrm{T} 
\left( z_+ - z_- \right) \mathscr{D}_{0+} \mathscr{D}_{0-} } \left[ \begin{array}{cc}
\dfrac{ c_\textrm{e}}{c_0} & -1 \\
-1 & \dfrac{ c_0}{c_\textrm{e}}
\end{array} \right],
\end{equation}
\begin{equation} \label{mz0+-}
\textbf{m}_z =  \frac{RT  \left( z_+ \mathscr{D}_{0-} + z_- \mathscr{D}_{0+}  \right) }{c_\textrm{T} 
\left\| \textbf{z} \right\|^2 \mathscr{D}_{0+} \mathscr{D}_{0-} } \left[ \begin{array}{c}
-1 \\
\dfrac{c_0}{c_\textrm{e}}
\end{array} \right],~~~~\textrm{and}
\end{equation}
\begin{equation}\label{Mzz0+-}
M_{zz} = - \tfrac{RT \left\| \textbf{z} \right\|^2 }{c_\textrm{T} z_+ z_- } \Big[  \tfrac{1}{\mathscr{D}_{+-} } +  \tfrac{\left( z_+ - z_- \right) \left( z_+^3  \mathscr{D}_{0-}+ z_-^3 \mathscr{D}_{0+} \right) c_0 }{\left\| \textbf{z} \right\|^4 \mathscr{D}_{0+}  \mathscr{D}_{0-} \nu c_\textrm{e}} \Big].
\end{equation}
\noindent Since the $\textbf{M}_Z^0$ matrix is invariant with respect to the choice of convective velocity, these expressions can be used in equation \eqref{OSM_neutral} or \eqref{neutOSMofJ}, or even alternative versions of equation \eqref{neutOSMofJ} in which the set of excess component fluxes relative to the mass-average velocity, $\vec{\textbf{j}}_Z^{\,0}$, is replaced with a set relative to some other velocity, $\vec{\textbf{j}}_Z^{\,0,\psi'}$.

Concentrated solution theory uses flux-explicit forms of the transport laws. Following Newman, we refer the excess fluxes to the solvent velocity (i.e., $\vec{N_0} / c_0$) by taking 
\begin{equation} \label{psiprimeZbinary}
\boldsymbol{\psi}'_Z = \frac{\textbf{Z}}{c_0^0} \left[ \begin{array}{c} 
1 \\ 
0 \\
0 
\end{array} \right]  = \frac{1}{c_0} \left[ \begin{array}{c} 
1 \\ 
0 \\
0 
\end{array} \right] 
\end{equation}
in equation \eqref{jZpsiprime}. This changes the dynamical variables to a set of electroneutral excess species fluxes
\begin{equation}\label{jZ0psiprime0+-}
\vec{\textbf{j}}_Z^{\,0,0} = \left[ \begin{array}{c}
0 \\
\vec{J}_\textrm{e}^{\,0,0} \\
\dfrac{ \vec{i} }{F\left\| \textbf{z} \right\|} 
\end{array} \right]
\end{equation}
whose first member is trivial, and whose second describes the excess salt flux relative to the solvent velocity. 

Since matrix $\textbf{M}_Z^0$ is built up of the sub-blocks defined by equations \eqref{Mnu0+-} through \eqref{Mzz0+-}, the transport matrix $\textbf{L}_Z^{0,0}$ for the flux-explicit constitutive formulation relative to the solvent velocity can be computed directly with equation \eqref{Lpsiprime_of_MZ0}, using $\boldsymbol{\psi}'_Z$ from equation \eqref{psiprimeZbinary}. The ionic conductivity is identified as the entry in the last row and column of $\textbf{L}_Z^{0,0}$ through equation \eqref{LZ0psiofDpsixipsi}, yielding
\begin{equation}\label{kappa0+-}
\kappa^0 = - \frac{c_\textrm{T} F^2 z_+ z_-}{RT} \left[ \frac{1}{\mathscr{D}_{+-}} + \frac{c_0 \left( z_+ - z_- \right)}{\nu c_\textrm{e} \left( z_+ \mathscr{D}_{0+} - z_- \mathscr{D}_{0-} \right)} \right]^{-1},
\end{equation}
whence the last column (or last row) of $\textbf{L}_Z^{0,0}$ reveals that
\begin{equation}\label{xipsi0+-}
\boldsymbol{\xi}^{0} = \left[ \begin{array}{c}
\xi_0^{0} \\
\xi_\textrm{e}^{0} 
\end{array} \right] = \left[ \begin{array}{c} 
0 \\
- \dfrac{\left( z_+ - z_- \right) \left( z_- \mathscr{D}_{0+} + z_+ \mathscr{D}_{0-} \right)}{\left\| \textbf{z} \right\| \nu \left( z_+ \mathscr{D}_{0+} - z_- \mathscr{D}_{0-} \right)}
\end{array} \right]
\end{equation}
are the component migration coefficients relative to the solvent velocity. Migration coefficients can be used to compute transference numbers with equation \eqref{tpsiprime}, yielding
\begin{equation}\label{tpsi0+-}
\textbf{t}^{0} = \left[ \begin{array}{c}
t_0^{0} \\
t_+^{0} \\
t_-^{0} 
\end{array} \right] = \left[ \begin{array}{c}
0 \\
\frac{z_+ \mathscr{D}_{0+} }{z_+ \mathscr{D}_{0+} -  z_- \mathscr{D}_{0-}} \\
- \frac{z_- \mathscr{D}_{0-} }{z_+ \mathscr{D}_{0+} -  z_- \mathscr{D}_{0-}}
\end{array} \right].
\end{equation}
\noindent The transference number $t_0^{0}$ vanishes necessarily because the solvent species is uncharged, conveniently matching the solvent component's null migration coefficient. Thus, for binary electrolytes, no information is lost by using transference numbers. 

A third flux-explicit transport property, $\mathfrak{D}$, is related to the Stefan--Maxwell coefficients by comparing the computed expression \eqref{Mnu0+-} with equation \eqref{component transport matrix}. This shows that
\begin{equation}\label{Dsalt0+-}
 \mathfrak{D} = \frac{ \left( z_+ - z_- \right) \mathscr{D}_{0+} \mathscr{D}_{0-} }{ z_+ \mathscr{D}_{0+} - z_- \mathscr{D}_{0-} }
\end{equation}
defines the thermodynamic diffusivity of salt in solvent. Given $\textbf{M}_\nu$ from equation \eqref{Mnu0+-} and that $\boldsymbol{\psi}'_\nu = [ 1/c_0 ~ 0 ]^\top $, Onsager component diffusivities relative to the solvent velocity, $\overline{\textbf{L}}_\nu^0$, are given by equation \eqref{barL_of_barM}, producing
\begin{equation}\label{Dnupsi0+-}
\overline{\textbf{L}}_\nu^{0} = \left[ \begin{array}{cc}
0 & 0 \\
0 & \dfrac{c_\textrm{e} \left( z_+ - z_- \right) \mathscr{D}_{0+} \mathscr{D}_{0-} }{\nu c_0 \left( z_+ \mathscr{D}_{0+} - z_- \mathscr{D}_{0-} \right) }.
\end{array} \right]
\end{equation}
\noindent Thus $\overline{\textbf{L}}_\nu^0$ is seen to be proportional to $\mathfrak{D}$, through a composition dependent prefactor.

Equations \eqref{kappa0+-}, \eqref{xipsi0+-}, and \eqref{Dsalt0+-} establish three macroscopic transport properties that parametrize equations \eqref{NewmanFluxLaws} and \eqref{NewmanMacInnes} in terms of Stefan--Maxwell diffusivities. Notably, equations \eqref{kappa0+-}, \eqref{tpsi0+-}, and \eqref{Dsalt0+-} respectively match the definition of ionic conductivity in Newman's equation 12.23, cation transference number relative to the solvent velocity in Newman's 12.11, and  thermodynamic diffusivity in Newman's 12.10 \cite{newman2004electrochemical}.

Identifying flux-explicit species laws and a MacInnes equation involving a single thermodynamic factor, with properties $\kappa^0$, $t_+^0$, and $\mathfrak{D}$ defined in terms of Stefan--Maxwell coefficients, completes Newman's development of concentrated solution theory. The structural properties laid out above make possible an opposite process, however, in which one writes flux laws a priori with the structure that kinematic and thermodynamic consistency necessitates. Once such laws have been formulated, they can be analyzed to find how the Onsager--Stefan--Maxwell matrix $\textbf{M}_Z^0$ depends on the posited set of flux-explicit properties. This route more naturally reflects how properties are typically measured, by fitting data to formulas derived from flux-explicit laws. It also develops the relationships needed to incorporate standard properties into numerical solvers that exploit the spectral structure of force-explicit transport formulations \cite{VanbruntIMA,VanbruntAIChE}.

For the binary electrolyte, $\textbf{i}_0$ (the first column of the identity matrix) sits in the nullspace of the symmetric Onsager component-diffusivity matrix relative to the solvent velocity, $\overline{\textbf{L}}_\nu^{0}$, which therefore has one nonzero entry (in the second row and second column),
\begin{equation}\label{positDnu0+-}
\overline{\textbf{L}}_\nu^{0} = \left[ \begin{array}{cc}
0 & 0 \\
0 & \mathscr{L}
\end{array} \right] = \left[ \begin{array}{cc}
0 & 0 \\
0 & \dfrac{c_\textrm{e} \mathfrak{D} }{\nu c_0 }
\end{array} \right],
\end{equation}
in which the single Onsager diffusivity of salt relative to the solvent velocity, $\mathscr{L}$, is expressed in terms of a thermodynamic diffusivity $\mathfrak{D}$ by applying inversion formula \eqref{barL_of_barM} to definition \eqref{component transport matrix}. Similarly, $\textbf{i}_0$ is orthogonal to $\boldsymbol{\xi}^{0}$, which therefore also must have just one nonzero entry, that is,
\begin{equation}\label{positxi0+-}
\boldsymbol{\xi}^{0} = \left[ \begin{array}{c}
0  \\
\xi, 
\end{array} \right]
\end{equation} 
where $\xi$ stands for the salt migration coefficient relative to the solvent velocity. This property relates to the cation transference number relative to solvent as
\begin{equation}\label{xie0+-t}
\xi = - \frac{\left( z_+ - z_- \right) \left[ \big( 1- t_+^{0} \big) z_+^2 - z_-^2 t_+^{0}  \right] }{\nu z_+ z_- \left\| \textbf{z} \right\| }
\end{equation}
through equation \eqref{tpsiprime}.

The known structures of $\overline{\textbf{L}}_\nu^{0}$ and $\boldsymbol{\xi}^{0}$ allow species flux laws relative to the solvent velocity to be written directly in the form of equation \eqref{NewmanFluxLaws}. Bringing in equations \eqref{binarystoich}, \eqref{tpsi0+-}, and \eqref{positDnu0+-}, one finds
\begin{align}
\vec{J}_0^{\,0} &= 0, \\
\vec{J}_+^{\,0} &= - \frac{\mathfrak{D} c_\textrm{T}^0 }{RT} \frac{\nu_+  c_\textrm{e}}{\nu c_0} \nabla \mu_\textrm{e}^0 + \frac{t_+^{0} \vec{i}}{Fz_+}, \label{Newman12.8} \\
\vec{J}_-^{\,0} &= - \frac{\mathfrak{D} c_\textrm{T}^0 }{RT} \frac{\nu_- c_\textrm{e}}{\nu c_0} \nabla \mu_\textrm{e}^0 + \frac{\big( 1- t_+^{0} \big) \vec{i}}{Fz_-} . \label{Newman12.9}
\end{align}
\noindent Note that equations \eqref{Newman12.8} and \eqref{Newman12.9} match Newman's equations 12.8 and 12.9, respectively \cite{newman2004electrochemical}.

Finally, posit an ionic conductivity $\kappa^0$ to write an electroneutral MacInnes equation in the form of \eqref{NewmanMacInnes}, as
\begin{equation}
\frac{\vec{i}}{\kappa^0} = - \nabla \Phi_z^0 - \frac{\left( z_+ - z_- \right) }{\nu F z_+ z_-} \left( \frac{z_+^2}{z_+^2+ z_-^2} - t_+^{0} \right) \nabla \mu_\textrm{e}^0,
\end{equation}
which depends on the salt--charge potential. Introducing a reference-electrode potential with equation \eqref{nablaPhiominus} gives
\begin{align}
\frac{\vec{i}}{\kappa^0} =& - \nabla \Phi^\ominus + \frac{s_0 \nabla \mu_0^0}{n_{\textrm{e}^-} F z_{\textrm{e}^-} } \nonumber \\ & + \frac{z_+ - z_-}{n_{\textrm{e}^-} F z_{\textrm{e}^-} \nu} \left[ t_+^{0} \frac{s_- }{z_+ } - \big( 1 - t_+^{0} \big) \frac{s_+ }{z_- } \right] \nabla \mu_\textrm{e}^0,
\end{align}
in harmony with Newman's equation 12.27 \cite{newman2004electrochemical}.

Given $\kappa^0$, as well as $\overline{\textbf{L}}_\nu^{0}$ and $\boldsymbol{\xi}^{0}$ matrices in the forms from equations \eqref{positDnu0+-} and \eqref{positxi0+-}, one can use relationships \eqref{MnuofL} and \eqref{Reconstruct M_Z} to construct the $\textbf{M}_{Z}^0$ matrix that would be needed to implement the electroneutral Onsager--Stefan--Maxwell equations. Performing the calculation, one finds
\begin{equation} \label{binaryM_nu}
\textbf{M}_\nu = \frac{RT }{c_\textrm{T}^0 \mathfrak{D} } \left[ \begin{array}{cc}
\frac{\nu c_\textrm{e}}{c_0} & -\nu \\
-\nu & \frac{\nu c_0}{c_\textrm{e}}
\end{array} \right],
\end{equation}
\begin{equation} \label{binarym_z}
\textbf{m}_z = \frac{\nu RT \xi }{ c_\textrm{T}^0 \mathfrak{D} } \left[ \begin{array}{c}
1 \\
- \tfrac{c_0}{c_\textrm{e}}  
\end{array} \right],
\end{equation}
and 
\begin{equation} \label{binaryM_zz}
M_{zz} = \frac{F^2 \left\| \textbf{z} \right\|^2 }{\kappa^0} + \frac{\nu RT c_0 \xi^2 }{c_\textrm{T}^0 c_\textrm{e} \mathfrak{D} }.
\end{equation}
One can also elucidate the Stefan--Maxwell coefficients that describe species interdiffusion, by inverting the congruence transformation in equation \eqref{MZ0block}: $\textbf{M}^0 = \textbf{Z}^{-1} \textbf{M}_Z^0 \textbf{Z}^{-\top}$. Off-diagonal entries of $\textbf{M}^0$ relate to $\mathscr{D}_{ij}$ values through equation \eqref{transport matrix}. Once $t_+^{0}$ has been introduced in favor of $\xi$ via equation \eqref{xie0+-t}, one finds
\begin{align}
\frac{1}{\mathscr{D}_{0+}} = & - \frac{ ( 1 - t_+^{0} ) \left( z_+ - z_- \right)}{z_- \mathfrak{D}  } , \nonumber \\
\frac{1}{\mathscr{D}_{0-}}= & \frac{ t_+^{0} \left( z_+ - z_- \right) }{ z_+ \mathfrak{D} } , \nonumber \\
\frac{1}{\mathscr{D}_{+-} }= & - \frac{F^2 z_+ z_- c_\textrm{T}^0 }{RT \kappa^0 } + \frac{t_+^0 ( 1 - t_+^{0} ) \left( z_+ - z_-\right)^2 c_0}{\nu \mathfrak{D} z_+ z_- c_\textrm{e} }, &
\end{align}
consistent with Newman's equations 14.1--14.3.

\subsection{The molten salt}
Pollard and Newman applied concentrated solution theory to the other possible liquid electrolyte comprising three species: a molten solution of two simple salts that share a common ion \cite{Newman-molten-salt1979}. From transport laws based on the Onsager--Stefan--Maxwell equations, they ultimately developed flux-explicit laws in the form of equation \eqref{NewmanFluxLaws} and a MacInnes equation like equation \eqref{NewmanMacInnes}. 

Here we apply the salt--charge framework to obtain transport laws consistent with Pollard and Newman's results. The flux laws we formulate will not precisely identify with the Pollard--Newman equations, however.  Whereas they used the common ion's velocity as the reference for convection, we instead use the velocity of one of the salts. This alternative convention creates component flux laws that involve a single chemical-potential gradient directly, rather than requiring additional use of the Gibbs--Duhem equation to eliminate a dependent driving force. 

Let the charges of the three species comprising the binary molten salt be $\textbf{z} = [z_{1}, z_{2}, z_{3}]^{\top}$. We assume that the third species is counter-charged to the first two, so that $\textrm{sgn} \left( z_1 \right) = \textrm{sgn} \left( z_2 \right)$ and $\textrm{sgn} \left( z_1 \right) = - \textrm{sgn} \left( z_3 \right)$. Again, the choice of products of the simple association equilibria is unique; we label the simple salts formed from these three species with indices corresponding to the similarly charged ion that distinguishes them: $\text{e}_{1}$ and $\text{e}_{2}$.  The stoichiometric coefficients of the ions within these salts are written in terms of the species charges and total formula-unit stoichiometries $\nu_{\text{e}_{1}}$ and $\nu_{\text{e}_{2}}$ as
\begin{align}
 & \nu_{1}^{\text{e}_{1}} = - \frac{\nu_{\text{e}_{1}} z_{3}}{z_{1} - z_{3}}, \:\:\:\:\:\: \nu_{3}^{\text{e}_{1}} =  \frac{\nu_{\text{e}_{1}}  z_{1}}{z_{1} - z_{3}}, \label{salt1} \\ 
& \nu_{2}^{\text{e}_{2}} = - \frac{\nu_{\text{e}_{2}} z_{3}}{z_{2} - z_{3}}, \:\:\:\:\:\: \nu_{3}^{\text{e}_{2}} =  \frac{\nu_{\text{e}_{2}}  z_{2}}{z_{2} - z_{3}}, \label{salt2}
\end{align}
so that $\nu_{\text{e}_{1}} = \nu_{1}^{\text{e}_{1}} + \nu_{3}^{\text{e}_{1}}$ and $\nu_{\text{e}_{2}} = \nu_{2}^{\text{e}_{2}} + \nu_{3}^{\text{e}_{2}}$. The matrix $\textbf{Z}$ that represents the salt--charge basis is unique up to ordering of the similarly charged ions. We choose to write it as
\begin{equation}\label{Z3ion}
\textbf{Z} =  \begin{bmatrix}
 \nu_{1}^{\text{e}_{1}} & 0 &  \nu_{3}^{\text{e}_{1}}  \\ 0 & \nu_{2}^{\text{e}_{2}}  & \nu_{3}^{\text{e}_{2}}  \\ \frac{z_{1}}{\| \textbf{z}\|} 
 &  \frac{z_{2}}{\| \textbf{z}\|} &  \frac{z_{3}}{\| \textbf{z}\|}
\end{bmatrix},
\end{equation} 
where $\left\| \textbf{z} \right\| = \sqrt{z_1^2+z_2^2+z_3^2}$, bearing in mind that the reactant stoichiometric coefficients from the simple association equilibria satisfy equations \eqref{salt1} and \eqref{salt2}. 

With the salt--charge transformation from equation \eqref{Z3ion}, the electroneutral species concentrations $\textbf{c}^{0}$ are cast as parametric functions of the component concentrations $c_{\textrm{e}_{1}}$ and $c_{\textrm{e}_{2}}$ through equation \eqref{c0def}, as 
\begin{equation}
\left[ \begin{array}{c}
c_1^0 \\
c_2^0 \\
c_3^0 
\end{array} \right] = \textbf{Z}^{\top} \left[ \begin{array}{c}
c_{\textrm{e}_{1}} \\
c_{\textrm{e}_{2}} \\
0
\end{array} \right] = \left[ \begin{array}{c}
- \frac{\nu_{\text{e}_1} z_3 }{z_1 - z_3} c_{\text{e}_1} \\
- \frac{\nu_{\text{e}_2} z_3 }{z_2 - z_3} c_{\text{e}_2} \\
\nu_{1}^{\textrm{e}_{1}} c_{\text{e}_{1}} + \nu_{2}^{\textrm{e}_{2}} c_{\text{e}_{2}} 
\end{array}
\right].
\end{equation}
\noindent The electroneutral component chemical potentials can then be written using equations \eqref{di_of_ynu} and \eqref{Xnu0def}. Together with the relation $\nu_{\textrm{e}_{2}} y_{\textrm{e}_{2}} = 1 - \nu_{\textrm{e}_{1}} y_{\textrm{e}_{1}} $, these show that 
\begin{align}
\vec{\nabla} \left[ \begin{array}{c}
\mu_{\textrm{e}_{1}}^{0} \\
\mu_{\textrm{e}_{2}}^0
\end{array} \right] 
=& \, RT\left[ \begin{array}{cc}
X_{\textrm{e}_{1}\textrm{e}_{1}}^0 & X_{\textrm{e}_{1}\textrm{e}_{2}}^0 \\
X_{\textrm{e}_{2}\textrm{e}_{2}}^0 & X_{\textrm{e}_{2} \textrm{e}_{2}}^0
\end{array} \right]\vec{\nabla} \left[ \begin{array}{c}
y_{\textrm{e}_{1}}\\
y_{\textrm{e}_{2}}
\end{array} \right] \nonumber \\
 =& \, RT \left[ \tfrac{z_3 \left( z_1 \nu_{\textrm{e}_1} y_{\textrm{e}_1} + z_2 \nu_{\textrm{e}_2} y_{\textrm{e}_2} \right)}{z_1 \left( z_3 - z_2 \right) \nu_{\textrm{e}_1} y_{\textrm{e}_1} + z_2 \left( z_3 - z_1 \right) \nu_{\textrm{e}_2} y_{\textrm{e}_2} } \right. \nonumber\\ 
& \left. + \tfrac{\nu_{\textrm{e}_1} y_{\textrm{e}_1} \nu_{\textrm{e}_2} y_{\textrm{e}_2} \Lambda^0_{\textrm{e}_1 \textrm{e}_1}}{\nu_{\textrm{e}_1}^2} \right] \left[ \begin{array}{c}
- \frac{1}{y_{\textrm{e}_1}} \\
\frac{1}{y_{\textrm{e}_2}}
\end{array} \right] \nu_{\textrm{e}_2} \nabla y_{\textrm{e}_2},
\end{align}
in which a single function $\Lambda_{\textrm{e}_{1}\textrm{e}_{1}} \left( y_{\textrm{e}_2} \right)$ parametrizes the composition gradient of either salt's mean molar activity coefficient within the solution. Observe that the ideal contribution to mixing free energy---the first term in the square brackets---depends in a rather subtle way on the salt fractions $y_{\textrm{e}_1}$ and $y_{\textrm{e}_2}$. 
 
Rather than repeating in its entirety the development illustrated for the binary electrolyte, we leverage the structural knowledge provided by our framework to abbreviate the process. Because the solution comprises three species, there are three flux-explicit transport properties $\mathfrak{D}$, $\xi$ and $\kappa^{0}$. Using the structure posited in equation \eqref{component transport matrix}, we may write
 \begin{equation}\label{Mnue1e2} 
 \textbf{M}_{\nu} = \frac{RT \nu_{\textrm{e}_{1}} \nu_{\textrm{e}_{2}}}{c_{\text{T}}^{0} \mathfrak{D}} \left[ \begin{array}{cc}
\dfrac{ c_{\textrm{e}_{2} }}{c_{\textrm{e}_{1}}} & -1\\
-1& \dfrac{ c_{\textrm{e}_{1} }}{c_{\textrm{e}_{2}}}
\end{array} \right].
 \end{equation}
\noindent In order to eliminate the chemical-potential gradient of salt 1 from the flux-explicit laws, it is convenient to choose $\boldsymbol{\psi}'$ such that $\boldsymbol{\psi}'_Z = \textbf{i}_{\textrm{e}_1} / c_{\textrm{e}_1}$, that is, to let the convective velocity be that of the first salt (i.e., $\vec{N}_{\textrm{e}_1}/c_{\textrm{e}_1}$). There is a single non-trivial migration coefficient $\xi$ relative to this reference velocity:
\begin{equation}
\boldsymbol{\xi}^{\textrm{e}_1}= \left[ \begin{array}{c}
0 \\ 
\xi 
\end{array} \right].
\end{equation}
\noindent Using equations \eqref{Reconstruct M_Z} then shows that
\begin{equation} \label{mze1e2}
\textbf{m}_z =  \frac{RT \nu_{\textrm{e}_{1}} \nu_{\textrm{e}_{2}} \xi}{c_{\text{T}}^{0} \mathfrak{D}} 
\left[ \begin{array}{c}
-1 \\
\dfrac{c_{\textrm{e}_{1}}}{c_{\textrm{e}_{2}}}
\end{array} \right]~~~~\textrm{and}
\end{equation}
\begin{equation}\label{Mzze1e2}
M_{zz} = \frac{F^{2} \|z\|^{2}}{\kappa^{0}} + \frac{RT \nu_{\textrm{e}_{1}} \nu_{\textrm{e}_{2}} \xi^{2}}{c_{\text{T}}^{0} \mathfrak{D}} \dfrac{c_{\textrm{e}_{1}}}{c_{\textrm{e}_{2}}}.
\end{equation}
\noindent Next, taking the transport matrix \eqref{transport matrix}, performing the congruence transformation \eqref{MZofM}, and comparing the result to equations \eqref{Mnue1e2}, \eqref{mze1e2}, and \eqref{Mzze1e2} allows identification of the three macroscopic transport properties in terms of Stefan--Maxwell coefficients. These are
\begin{align}
\mathfrak{D} = & \frac{c^0_{3} (z_{1} - z_{3}) (z_{2} -z_{3})  }{\left( \frac{c^0_{3} z_{3}^{2}}{\mathscr{D}_{12}} +\frac{c^0_{2} z_{2}^{2}}{\mathscr{D}_{13}}  + \frac{c^0_{1} z_{1}^{2}}{\mathscr{D}_{23}} \right )} , \label{mathfrakDmolten} \\
\xi 
 \cdot \tfrac{\nu_{\text{e}_{1}} \nu_{\text{e}_{2}}  c^0_{1} c^0_{3} z_{3}^{2}}{\mathfrak{D} \nu_{1}^{\textrm{e}_{1}} } =& \tfrac{z_{3}^{2} c_{3} (c^0_{1} z_{2} - c^0_{2} z_{1})}{\mathscr{D}_{12}\|z\|^{2}}   -\tfrac{z_{2} c^0_{1} c^0_{2} (z_{1}^{2} +z_{3}^{2}) + ({c^0_{2}})^{2} z_{1} z_{2}^{2} }{\mathscr{D}_{13}\|z\|^{2}}  \nonumber
 \\ 
& + \tfrac{({c^0_{1}})^{2} z_{1}^{2}z_{2} + c^0_{1} c^0_{2} z_{1} (z^{2}_{2} + z_{3}^{2})}{\mathscr{D}_{23}\|z\|^{2}}, \\
\kappa^0 = & \frac{ F^{2} c_{\textrm{T}}^{0}  \left ( \frac{z_{1}^{2} c_{1}^0}{\mathscr{D}_{23}} + \frac{z_{2}^{2} c_{2}^0}{\mathscr{D}_{13}}  + \frac{z_{3}^{2} c_{3}^0}{\mathscr{D}_{12}} \right )}{ RT \left ( \frac{c_{1}^0}{\mathscr{D}_{12} \mathscr{D}_{13}} + \frac{c_{2}^0}{\mathscr{D}_{12} \mathscr{D}_{23}}  + \frac{c_{3}^0}{\mathscr{D}_{13} \mathscr{D}_{23}} \right ) }. \label{Kappamolten}
\end{align}
\noindent Here the conductivity is identical to Pollard and Newman's equation 33 \cite{Newman-molten-salt1979}. The diffusivity defined in equation \eqref{mathfrakDmolten} relates to the diffusion coefficient identified in their equation 9 \cite{Newman-molten-salt1979}, which we will label $\mathscr{D}$, as
\begin{equation}
\mathfrak{D} = \mathscr{D} \frac{c_{\text{T}}^{0}}{c_{3}^0} \frac{\nu_{\text{e}_{1}}\nu_{\text{e}_{2}}}{\nu^{\text{e}_{1}}_{1}\nu^{\text{e}_{2}}_{1}}.
\end{equation}  
\noindent Although the prefactor that relates our $\mathfrak{D}$ to Pollard and Newman's $\mathscr{D}$ is composition-dependent, the factor only involves the particle fraction of the common ion, which is always of order unity.

It is equally straightforward to invert this process. Similar to the binary-electrolyte analysis, entries of $\textbf{M}^0$ may be related to the original Stefan--Maxwell diffusivities $\mathscr{D}_{ij}$ by inverting the congruence transformation in equation \eqref{MZ0block} and examining the off-diagonal elements in light of equation \eqref{neutral_transport_matrix}. This comparison shows that the diffusivities relate to the macroscopic properties as
\begin{align}
\tfrac{1}{\mathscr{D}_{12}} = & \tfrac{ \left[ c_{1}^{0} (z_{1}^{2} + z_{3}^{2} ) + c_{2}^{0} z_{1} z_{2} \right] \left[ c_{2}^{0} (z_{2}^{2}+z_{3}^{2}) +c_{1}^{0} z_{1} z_{2} \right] \nu_{\textrm{e}_{1}}  \nu_{\textrm{e}_{2}}}{\mathfrak{D} c_{1}^{0} c_{2}^{0} \nu_{1}^{\textrm{e}_{1}}  \nu_{2}^{\textrm{e}_{2}} \| \textbf{z}\|^{4}}
  \nonumber \\
 & -  \tfrac{\xi \nu_{\textrm{e}_{1}}\nu_{\textrm{e}_{2}} [ c_{1}^{0}  z_{1}(z_{1}^{2}-z_{2}^{2}+z_{3}^{2})  -c_{2}^{0} z_{2} ( -z_{1}^{2} + z_{2}^{2} +z_{3}^{2})] }{ \mathfrak{D} \|\textbf{z}\|^{3} c_{2}^{0} \nu^{\textrm{e}_1}_{1} }  \nonumber \\
 & - \left(\tfrac{F^{2} c_{\text{T}}^{0}}{RT\kappa^0 } +\tfrac{\nu_{\textrm{e}_{1}} \nu_{\textrm{e}_{2}} \xi^{2}}{ \mathfrak{D}} \tfrac{c_{\textrm{e}_{1}}}{c_{\textrm{e}_{2}}} \right) \tfrac{z_{1} z_{2}}{\|\textbf{z}\|^{2}} \nonumber
\\
  \tfrac{1}{\mathscr{D}_{13}} = &- \tfrac{z_{3} (c_{1}^{0} z_{2}- c_{2}^{0} z_{1} )\left[ c_{2}^{0} (z_{2}^{2}+z_{3}^{2}) +c_{1}^{0} z_{1} z_{2} \right] \nu_{\textrm{e}_{1}} \nu_{\textrm{e}_{2}} }{\mathfrak{D} c_{1}^{0} c_{2}^{0}  \nu_{1}^{\textrm{e}_{1}} \nu_{2}^{\textrm{e}_{2}} \| \textbf{z}\|^{4}}
  \nonumber \\
 & + \tfrac{\xi z_{3} \nu_{\textrm{e}_{1}}\nu_{\textrm{e}_{2}}  \left[ c_{2}^{0}  (-z_{1}^{2}+z_{2}^{2} + z_{3}^{2} )+2 c_{1}^{0}   z_{1} z_{2} \right]}{\mathfrak{D} c_{2}^{0} \nu_{1}^{\textrm{e}_{1}} \|\textbf{z} \|^{3}} \nonumber \\ 
 & - \left( \tfrac{F^{2} c_{\text{T}}^{0}}{RT\kappa^0 } + \tfrac{\nu_{\textrm{e}_{1}} \nu_{\textrm{e}_{2}} \xi^{2}}{ \mathfrak{D}} \tfrac{c_{\textrm{e}_{1}}}{c_{\textrm{e}_{2}}} \right ) \tfrac{z_{1} z_{3}}{\|\textbf{z}\|^{2}} \nonumber \\
\tfrac{1}{\mathscr{D}_{23}} = &  -\tfrac{z_{3} (c_{2}^{0} z_{1}- c_{1}^{0} z_{2} )\left[ c_{2}^{0}  (z_{1}^{2}+z_{3}^{2}) +c_{2}^{0}  z_{1} z_{2} \right] \nu_{\textrm{e}_{1}}\nu_{\textrm{e}_{2}} }{ \mathfrak{D} c_{1}^{0} c_{2}^{0} \nu_{1}^{\textrm{e}_{1}}  \nu_{2}^{\textrm{e}_{2}} \| \textbf{z}\|^{4}}
  \nonumber \\
 & + \tfrac{\xi z_{3} \nu_{\textrm{e}_{1}}\nu_{\textrm{e}_{2}} \left[ c_{1}^{0} (z_{1}^{2}-z_{2}^{2} + z_{3}^{2} )+2 c_{2}^{0}  z_{1} z_{2} \right]}{\mathfrak{D} c_{2}^{0}  \nu_{1}^{\textrm{e}_{1}} \|\textbf{z} \|^{3}} \nonumber \\
 & - \left( \tfrac{F^{2} c_{\text{T}}^{0}}{RT\kappa^0 } +\tfrac{\nu_{\textrm{e}_{1}} \nu_{\textrm{e}_{2}} \xi^{2}}{ \mathfrak{D}} \tfrac{c_{\textrm{e}_{1}}}{c_{\textrm{e}_{2}}} \right )\tfrac{z_{2} z_{3}}{\|\textbf{z}\|^{2}}. 
\end{align}
\noindent Pollard and Newman do not write down these relationships, presumably because of their algebraic complexity. Stating explicit forms for the Stefan--Maxwell coefficients in terms of flux-explicit properties becomes increasingly cumbersome as the number of species (and the number of charged species) in a system increases. 

With our choice of convective velocity, the matrix $\overline{\textbf{L}}_{\nu}^{\,\textrm{e}_1}$ also affords a single non-trivial entry,
\begin{equation}
\overline{\textbf{L}}_{\nu}^{\,\textrm{e}_1} = 
\left[ \begin{array}{cc}
0 & 0  \\
0 & \frac{\mathfrak{D}}{ \nu_{\textrm{e}_{1}}\nu_{\textrm{e}_{2}}} \frac{ c_{\textrm{e}_{2}}}{ c_{\textrm{e}_{1}}}
\end{array} \right].
\end{equation}
\noindent After deriving transference numbers $t_{1}^{\textrm{e}_1}$, $t_{2}^{\textrm{e}_1}$, and $t_{3}^{\textrm{e}_1}$ relative to salt 1 with equation \eqref{tpsiprime}, the laws governing the species excess fluxes are found to be
\begin{align}
\vec{J}_1^{\,\textrm{e}_1} &=  \frac{t_{1}^{\textrm{e}_1} \vec{i}}{F z_{1}}, \\
 \vec{J}_{2}^{\,\textrm{e}_1} &=  -  \frac{ c_{\textrm{T}}^{0} \mathfrak{D}}{ RT} \frac{\nu^{\textrm{e}_{1}}_{1} c_{\textrm{e}_{2}}}{  \nu_{\textrm{e}_{1}}\nu_{\textrm{e}_{2}} c_{\textrm{e}_{1}}}\nabla \mu_{\textrm{e}_{2}}^0 + \frac{t_2^{\textrm{e}_1} \vec{i}}{Fz_2} ,
 \\
\vec{J}_3^{\,\textrm{e}_1} &= -  \frac{ c_{\textrm{T}}^{0} \mathfrak{D}}{ RT} \frac{\nu^{\textrm{e}_{2}}_{2} c_{\textrm{e}_{2}}}{  \nu_{\textrm{e}_{1}}\nu_{\textrm{e}_{2}} c_{\textrm{e}_{1}}}\nabla \mu_{\textrm{e}_{2}}^0  + \frac{ t_{3}^{\textrm{e}_1}  \vec{i}}{Fz_3}  . \label{Newman12.9b}
\end{align}
\noindent Interestingly, the excess flux of ion 1 relative to the velocity of salt 1 does not vanish; it can be driven by migration, but not by diffusion. Equation \eqref{tpsiprime} shows that all three transference numbers here are parametric functions of the single migration coefficient $\xi$, through 
\begin{equation}\label{t_of_xi_3ion}
\left[ \begin{array}{c}
t_1^{\textrm{e}_1} \\
t_2^{\textrm{e}_1} \\
t_3^{\textrm{e}_1} 
\end{array}
\right]
= \frac{\diag \left( \textbf{z} \right) \textbf{Z}^\top}{\left\| \textbf{z} \right\| }\left[ \begin{array}{c}
0 \\
\xi \\
1
\end{array}
\right].
\end{equation}
\noindent Thus, as well as satisfying the relation $t_{1}^{\textrm{e}_1} + t_{2}^{\textrm{e}_1} + t_{3}^{\textrm{e}_1} = 1$ that follows from the last row of this matrix equation, the species transference numbers relative to salt 1 are also constrained such that  $(z_2^2 + z_3^2) t_1^{\textrm{e}_1} + z_1^2 ( t_2^{\textrm{e}_1}+ t_3^{\textrm{e}_1}) = 0$, which follows from the first row. 

The reference velocity under $\boldsymbol{\xi}^{\textrm{e}_1}$ can be changed to that of the common ion by moving through the mass-average velocity with transformation \eqref{xipsiprime}. Then one can derive the associated transference numbers, denoted $t_{i}^{3}$, with equation \eqref{tpsiprime} to find Pollard and Newman's expression
\begin{equation}
t_{2}^{3} = \frac{  \frac{z_{2}}{\mathscr{D}_{13}} - \frac{z_{3}}{\mathscr{D}_{12}} }{\left ( \frac{z_{2}}{\mathscr{D}_{13}} - \frac{z_{3}}{\mathscr{D}_{12}} \right )  + \frac{z_{1} c^0_{1}}{z_{2} c^0_{2}} \left( \frac{z_{1}}{\mathscr{D}_{23}} - \frac{z_{3}}{\mathscr{D}_{12}} \right ) },
\end{equation}
which appears in their equation 10 \cite{Newman-molten-salt1979}. With a migration coefficient relative to salt 1 and the reference-electrode potential from equation \eqref{nablaPhiominus}, one finds a concise MacInnes equation,
\begin{align}
\frac{\vec{i}}{\kappa^0} =& - \nabla \Phi^\ominus + \frac{s_{\textrm{e}_{1}} \nabla \mu_{\textrm{e}_{1}}^0}{n_{\textrm{e}^-} F z_{\textrm{e}^-} }  + \left( \frac{s_{\textrm{e}_{2}} }{n_{\textrm{e}^-} F z_{\textrm{e}^-} } - \frac{\kappa^{0} \xi}{F \| \textbf{z} \|} \right) \nabla \mu_{\textrm{e}_{2}}^0,
\end{align}
in agreement with Pollard and Newman's equation 32 \cite{Newman-molten-salt1979}. 
 
 \subsection{The cosolvent electrolyte}
For a final illustration of our method of developing flux laws for concentrated multicomponent electrolytes, we consider a four-species liquid solution comprising two uncharged species, which we call a \emph{cosolvent}, and a single salt. Electrolytic solutions of this type are used in lithium-ion batteries, which typically rely on blends of linear and cyclic carbonates to dissolve simple lithium salts \cite{Xu2004}. We let subscripts $+$ and $-$ describe the cation and anion, respectively, a subscript $\textrm{e}$, the salt they associate to form, and let subscripts $0$ and $\text{o}$ stand for the two solvents. As was the case in the binary-electrolyte example, the stoichiometric coefficients for the salt satisfy equation \eqref{binarystoich}, where $\nu$ is the total salt stoichiometry.
 
A salt--charge basis for the cosolvent electrolyte is embodied by the transformation matrix
\begin{equation}
\textbf{Z} = \ \begin{bmatrix}
1 & 0 & 0 & 0 \\ \  0 & 1& 0 & 0 \\  0 & 0 & -\frac{\nu z_{-}}{z_{+}-z_{-}} &\frac{\nu z_{+}}{z_{+} - z_{-}} \\
0 & 0 & \frac{z_{+}}{\|\textbf{z}\|} & \frac{z_{-}}{\|\textbf{z} \|} 
\end{bmatrix}.
\end{equation}
\noindent The species concentrations in the neutral state can then be written as
\begin{equation}
\left[ \begin{array}{c}
c_0^0 \\
c_{\textrm{o}}^{0} \\
c_+^0 \\
c_-^0 
\end{array} \right] = \textbf{Z} \left[ \begin{array}{c}
c_0 \\ 
c_{\textrm{o}}\\
c_\textrm{e} \\
0
\end{array} \right] = \left[ \begin{array}{c}
c_0 \\
c_{\textrm{o}}\\
- \frac{z_- \nu }{z_+ - z_-} \cdot c_\textrm{e}  \\
\frac{z_+ \nu }{z_+ - z_-} \cdot c_\textrm{e}
\end{array}
\right],
\end{equation} 
parametric functions of the three component concentrations $c_0$, $c_\textrm{o}$, and $c_\textrm{e}$.

The nonideal parts of the electroneutral component chemical potentials are parametrised by a symmetric $2 \times 2$ block matrix
\begin{equation}
\boldsymbol{\Lambda} =   \left[ \begin{array}{cc}
\Lambda_{0 0} & \Lambda_{0 \text{o}} \\
\Lambda_{0 \text{o}} & \Lambda_{\text{o} \text{o}},
\end{array} \right],
\end{equation}
whose three independent entries are functions of only two of the three component fractions $y_0$, $y_\textrm{o}$, and $y_\textrm{e}$. Following equation \eqref{gradmunu0ofLambda}, the gradients of the electroneutral component chemical potentials are  expanded as
\begin{align}
  \nabla \left[ \begin{array}{c}
\mu_{0}^{0} \\
\mu_{\textrm{o}}^0 \\
\mu_{\text{e}}^{0} 
\end{array} \right]  =\,& RT  \nabla \ln \textbf{y}_\nu 
\nonumber \\
 & + RT (\textbf{I}_{3} - \boldsymbol{\nu} \textbf{y}^{\top}_{\nu} ) \left[ \begin{array}{cc}
\boldsymbol{\Lambda} & \textbf{o} \\
\textbf{o}^\top & 0
\end{array} \right ]
\nabla 
\textbf{y}_\nu ,
\end{align}
where in this case $\boldsymbol{\nu} = [1, 1, \nu]^\top$ and $\textbf{y}_{\nu} = [y_{0}, y_{\textrm{o}}, y_{\textrm{e}}]^\top $. 

To derive flux explicit laws, we again use the ansatz \eqref{component transport matrix} to write
\begin{equation}
\frac{c_{\textrm{T}}^{0} \textbf{M}_{\nu}}{RT} = \begin{bmatrix}
   \tfrac{c_{\text{o}}}{c_{0} \mathfrak{D}_{0\text{o}}} + \tfrac{\nu c_{\text{e}}}{c_{0} \mathfrak{D}_{0\textrm{e}}} & - \tfrac{1}{\mathfrak{D}_{0\text{o}}} & - \tfrac{\nu}{ \mathfrak{D}_{0\textrm{e}}}  \\   
   - \tfrac{1}{\mathfrak{D}_{0\text{o}}} &   \tfrac{c_{0}}{c_{\text{o}} \mathfrak{D}_{0\text{o}}}  + \tfrac{ \nu c_{\text{e}}}{c_{\textrm{o}} \mathfrak{D}_{\textrm{oe}}} & -\tfrac{\nu}{\mathfrak{D}_{\textrm{oe}}} \\  
   -\tfrac{\nu}{ \mathfrak{D}_{0\textrm{e}}}  & -\tfrac{\nu}{\mathfrak{D}_{\textrm{oe}}} & \tfrac{\nu c_{0}}{c_{\text{e}} \mathfrak{D}_{0\text{e}}} + \tfrac{ \nu c_{\text{o}} }{c_{\text{e}}\mathfrak{D}_{\textrm{oe}}}
\end{bmatrix}.
\end{equation}
These entries straightforwardly relate to the Stefan--Maxwell coefficients through
\begin{align}
 & \mathfrak{D}_{0\text{o}} = \mathscr{D}_{0\text{o}}, \\
 & \mathfrak{D}_{0\textrm{e}} = \frac{(z_{+}-z_{-})\mathscr{D}_{0+} \mathscr{D}_{0-} }{\mathscr{D}_{0+}z_{+} - \mathscr{D}_{0-}z_{-}}, \\
& \mathfrak{D}_{\textrm{oe}} = \frac{(z_{+}-z_{-})\mathscr{D}_{\text{o}+} \mathscr{D}_{\text{o}-}  }{\mathscr{D}_{\text{o}+}z_{+} - \mathscr{D}_{\text{o}-}z_{-}}. 
\end{align}
\noindent Observe that the second two definitions take forms similar to equation \eqref{Dsalt0+-}, despite the addition of a solvent.

Taking the reference velocity to be that of the first solvent, i.e., $\boldsymbol{\psi}' = \textbf{i}_0 / c_0$, the diffusivity matrix and migration coefficients assume the forms
 \begin{equation}
\overline{\textbf{L}}^{0}_{\nu}  =
 \begin{bmatrix}
 0 & 0 & 0 \\   0 & \mathscr{L}_{\textrm{o}} & \mathscr{L}_{\times} \\  0 & \mathscr{L}_{\times} & \mathscr{L}_{\textrm{e}}
\end{bmatrix}, \:\:\:\:\:\: \boldsymbol{\xi}^0 = \left [ \begin{array}{c}
0 \\ \xi_{\textrm{o}} \\ \xi_{\textrm{e}}
\end{array}
\right].
 \end{equation}
\noindent The terms appearing in the diffusion matrix are understood as follows. $\mathscr{L}_{\textrm{o}}$ denotes the Onsager diffusivity of the second solvent relative to the velocity of the first, $\mathscr{L}_{\times}$ the cross-diffusivity between the second solvent and the salt in a frame moving with the first solvent, and $\mathscr{L}_{\textrm{e}}$ the diffusivity of the salt relative to the first solvent's velocity. Each of these parameters is independent, and the $2 \times 2$ sub-block of Onsager diffusivities is positive definite. The set of migration coefficients depends on two independent parameters $\xi_{\textrm{o}}$ and $\xi_{\textrm{e}}$. A distinction from the previous cases is that a neutral species---the second solvent---may carry current. Thus transference numbers as typically understood do not suffice to write complete flux-explicit transport laws for this system. 

By relation \eqref{barL_of_barM}, the Onsager diffusivities that make up $\overline{\textbf{L}}^{0}_{\nu}$ can be understood in terms of component diffusivities, as
\begin{align}
&  \mathscr{L}_{\textrm{o}} = \frac{c_{\text{o}}\mathfrak{D}_{0\text{o}}(\mathfrak{D}_{\text{oe}}c_{0} + \mathfrak{D}_{0\text{e}} c_{\text{o}})}{c_{0} (\mathfrak{D}_{\text{oe}}c_{0} + \mathfrak{D}_{0\text{o}}\nu c_{\text{e}} + \mathfrak{D}_{0\text{e}}c_{\text{o}})} \\
 & \mathscr{L}_{\textrm{e}} = \frac{c_{\text{e}}\mathfrak{D}_{0\text{e}}(\mathfrak{D}_{\text{oe}}c_{0} + \mathfrak{D}_{0\text{o}}\nu c_{\text{e}})}{c_{0} (\mathfrak{D}_{\text{oe}}c_{0} + \mathfrak{D}_{0\text{o}}\nu c_{\text{e}} + \mathfrak{D}_{0\text{e}}c_{\text{o}})} \\
 & \mathscr{L}_{\times} = \frac{\mathfrak{D}_{0\textrm{o}}\mathfrak{D}_{0\textrm{e}}c_{\text{o}}c_{\text{e}}}{c_{0}(\mathfrak{D}_{\textrm{oe}} c_{0} + \nu \mathfrak{D}_{0\textrm{o}}c_{\textrm{e}} + \mathfrak{D}_{0\textrm{e}}c_{\textrm{o}})}.
\end{align}
\noindent Each of these expressions agrees within a prefactor with the thermodynamic oxygen diffusivity, the thermodynamic electrolyte diffusivity, and the cross diffusivity proposed to describe lithium/air-battery electrolytes by Monroe \cite{4componentLiO2}.
 
The benefits of relying on two migration coefficients and directly writing flux laws in the salt--charge basis are made clearest by this example. Equation 
 \eqref{ComponentFick_psiprime} reads for the two-solvent case as 
\begin{align} \label{2solventfluxexplicit1}
& \vec{J}_{0}^{\,0} = \vec{0}, \\ \label{2solventfluxexplicit2}
& \vec{J}_{\textrm{o}}^{\,0} = -\mathscr{L}_{\textrm{o}} \frac{c_{\text{T}}^{0}\nabla \mu_{\textrm{o}}}{RT}-\mathscr{L}_{\times} \frac{c_{\text{T}}^{0} \nabla \mu_{\textrm{e}}}{RT}+ \xi_{\textrm{o}} \frac{\vec{i}}{F \|z\|}  \\ \label{2solventfluxexplicit3}
& \vec{J}_{\textrm{e}}^{\,0} =   -\mathscr{L}_{\times} \frac{c_{\text{T}}^{0} \nabla \mu_{\textrm{o}}}{RT}-\mathscr{L}_{\textrm{e}}  \frac{ c_{\text{T}}^{0}\nabla \mu_{\textrm{e}}}{RT} + \xi_{\textrm{e}} \frac{\vec{i}}{F \|z\|}. 
\end{align}
\noindent Equations \eqref{2solventfluxexplicit1}-\eqref{2solventfluxexplicit3} are equivalent to equations 14-16 in the paper by Monroe \cite{4componentLiO2}. 
These laws combine with the MacInnes equation for a reference electrode, written through equations \eqref{NewmanMacInnes} and \eqref{nablaPhiominus} as 
\begin{equation} \label{2solventMacInnes}
\vec{i} = - \kappa^{0} \nabla \Phi^\ominus  + \frac{\kappa^0 \textbf{s}_{\nu}^\top \nabla \boldsymbol{\mu}_\nu^0 }{n_{\textrm{e}^-} F z_{\textrm{e}^-}  } -   \frac{\kappa^0 {\boldsymbol{\xi}^0}^\top  \nabla \boldsymbol{\mu}_\nu^0}{F\left\| \textbf{z} \right\| }, 
\end{equation}
to offer a complete description of isothermal, isobaric mass transport in the cosolvent electrolyte. 

One could proceed as in the previous examples to elucidate how the migration coefficients, diffusivities, and conductivity relate to the Stefan--Maxwell coefficients. This would show, for example, that the migration coefficient $\xi_k$ is expected to be proportional to component concentration $c_{\nu,k}$, as observed by Monroe \cite{4componentLiO2}. We refrain from completing this exercise here because of the meteoric rise in complexity of the intermediate equations. On account of the matrix inversion involved when moving between force- and flux-explicit forms, writing out entries within both the thermodynamic- and transport-property matrices becomes increasingly untenable as the number of species increases. Nevertheless, this process was implemented and found to produce relationships among parameters consistent with those reported in detail elsewhere \cite{4componentLiO2}. 

Matrix forms that connect flux- and force-explicit representations of transport laws are compact and generally useful. The operations described in prior sections are readily implemented with symbolic-manipulation software for systems containing large numbers of species. Perhaps most important is that the structures of the coefficient matrices within the electroneutral transport laws are very general. As shown in these examples, structural knowledge can be leveraged to create consistent sets of thermodynamic and transport constitutive laws a priori in either the Stefan--Maxwell or Fick representation, for electrolytes involving any number of components. 

\section{Numerical implementation: Hull cell}

Recently, the authors presented finite-element methods to simulate Onsager--Stefan--Maxwell transport of multiple uncharged species \cite{VanbruntIMA,VanbruntAIChE}. The efficacy of these approaches generally relies on the spectral structure of the coefficient matrix $\textbf{M}$. A significant advantage of the electroneutral Onsager--Stefan--Maxwell equations developed here is that they can be solved easily within this numerical scheme. Indeed, it was discussed in sections \ref{sec:GEs_saltcharge} and \ref{subsec:Elec_and_Dyn} how Sylvester's law of inertia guarantees that the spectral structure of electrolyte-transport laws remains essentially intact, in that the electroneutral transport matrix over the salt--charge basis, $\textbf{M}_{Z}^{0}$, is symmetric positive semidefinite with one null eigenvalue corresponding to the column $\textbf{c}^{0}_{Z}$. 
  
In the prior papers \cite{VanbruntIMA} and \cite{VanbruntAIChE}, the numerical challenge posed by singular transport matrices was resolved with a matrix augmentation. In the present context, for a given reference velocity in the component space, $\boldsymbol{\psi}'_Z$, we solve the augmented electroneutral Onsager--Stefan--Maxwell equations as
\begin{equation}
-\vec{\nabla} \mu_{Z}  = \textbf{M}_{Z}^{0,\psi'} \left( \gamma \right) \vec{\textbf{n}}_{Z},
\end{equation}
where $\textbf{M}_{Z}^{0,\psi'_Z} \left( \gamma \right) = \textbf{M}_{Z}^0 + \gamma \boldsymbol{\psi}'_{Z} \boldsymbol{\psi}'^{\top}_{Z}$ proves to be a symmetric positive definite matrix whenever $\gamma \ne 0$ (see reference \cite{VanbruntIMA}). These equations are coupled to the component balances \eqref{cnubalances} and electroneutral charge balance \eqref{Kirchhoff_node}.

We demonstrate how the framework created above can facilitate numerical analysis by solving for the steady-state current distribution in a symmetric electrode Hull cell containing a nonideal binary electrolytic solution, from which cations are plated and stripped under potentiostatic conditions, assuming that the electrode reactions have very low overpotential (fast kinetics). To show how thermodynamic consistency and system nonideality can be brought into simulations in a meaningful way, we take the electrolyte to be lithium hexafluorophosphate (LiPF$_{6}$) in ethyl--methyl carbonate (EMC)---a formulation used in lithium-ion batteries---whose properties were measured as functions of composition by Wang et al.\ \cite{AndrewLIPF6, AndrewMRI}. This electrolyte has total stoichiometry $\nu = 2$ and the species charges are $\textbf{z} = \left[ z_0, z_+, z_- \right]^\top = \left[ 0, 1, -1 \right]^\top$; the transformation matrix to the salt-charge basis we employ is
\begin{equation}\label{Zlipf6}
\textbf{Z} =  \begin{bmatrix}
1 & 0 & 0 \\ 0 & 1 &  1 \\
0 & \frac{1}{\sqrt{2}} & -\frac{1}{\sqrt{2}}
\end{bmatrix},
\end{equation}
in which the first two rows correspond to the simple association equilibria
\begin{equation}
\text{EMC} \rightleftharpoons \text{EMC} \:\:\:\:\:\: \text{and} \:\:\:\:\:\: \text{Li}^{+} + \text{PF}_{6}^{-} \rightleftharpoons \text{LiPF}_{6}.
\end{equation}
\noindent Wang's correlations for the composition dependences of ionic conductivity $\kappa$, Fickian diffusivity $D$, and Darken factor $\chi$ are expressed in terms of the salt fraction $y_\textrm{e}$ and solvent fraction $y_0 = 1 - \nu y_\textrm{e}$ in Table \ref{tab:LiPF6EMCprops}.
\begin{table}
\centering
\begin{tabular}{ |p{0.8cm}|p{5.5cm}|p{1cm}| }
\hline
\multicolumn{3}{|c|}{Properties} \\
\hline
Term & Function & Units \\
\hline
\hline
$\kappa$ & $\left ( 48.93 y_{\text{e}}^{3/2} - 284.8 y_{\text{e}}^{5/2} 817.8 y_{\text{e}}^{4} \right )^{2}$ & $\text{S m}^{-1}$ \\
$D$ & $\left( 4.998 - 29.96y_{\text{e}} + 53.78 y_{\text{e}}^{2} \right) \times 10^{-10} $  & $\text{m}^{2} \text{s}^{-1}$ \\
$\chi$ & $ 1 -18.38 y_{\text{e}}^{\frac{1}{2}} + 155.3y_{\text{e}} - 450.6y_{\text{e}}^{\frac{3}{2}} + 1506 y_{\text{e}}^{\frac{5}{2}}$ & $-$ \\
$t_{+}^{0}$  &$0.4107  -1.487 y_{\text{e}} + 2.547 y_{\text{e}}^{2}$ & $- $ \\
$\rho$ & $1007.1+ 10^{5} \times \big ( 0.0180 y_{\text{e}} - 0.1946 y_{\text{e}}^2 + 1.960 y_{\text{e}}^{3} - 7.008 y_{\text{e}}^{4}  + 8.004 y_{\text{e}}^{5} \big ) $  & $\text{g} \: \text{L}^{-1}$ \\
 \hline
\end{tabular}
\caption{Phenomenological property data necessary to specify the transport matrix $\textbf{M}$ and thermodynamic factor $\boldsymbol{\Lambda}$ for solutions of lithium hexafluorophosphate in ethyl--methyl carbonate at room temperature, derived from data reported by Wang et al.\ \cite{AndrewLIPF6, AndrewMRI}. \label{tab:LiPF6EMCprops}} 
\end{table} 
Note that $D$ and $\chi$ determine the properties $\Lambda_{00}$ and $\mathfrak{D}$ defined in section \ref{subsec:binaryelectrolyte} through
\begin{align} \label{LiPF6DandChirelations}
& \mathfrak{D} = \frac{D}{\chi}, \\
&  \Lambda_{00} = \frac{\chi - 1}{\nu y_{\textrm{e}} y_{0}} .
\end{align}
\noindent The migration coefficient $\xi$ is determined by the cation transference number relative to the solvent velocity $t_+^0$ through equation \eqref{xie0+-t}, which reduces to
\begin{equation} \label{LiPF6xirelation}
\xi =  \frac{1- 2 t_{+}^{0}}{\sqrt{2}}
\end{equation}
for this electrolyte system. Finally, it should be noted that the composition dependence of $\rho$ appearing in Table \ref{tab:LiPF6EMCprops} was not reported by Wang et al., but instead derives from a least-squares fit of the densitometric data reported in the supplementary information of those papers as a function of mass fraction \cite{AndrewLIPF6, AndrewMRI}. For this system, $\rho \left( y_\textrm{e} \right)$ acts as the volumetric equation of state. When using density to compute properties relevant to concentration from $\rho$, the molar mass of EMC was taken to be $\overline{m}_{0} = 104.105$ $\text{g}\, \text{mol}^{-1}$, and that of LiPF$_{6}$, $\overline{m}_{\text{e}} = 151.905$ $\text{g}\, \text{mol}^{-1}$. After inferring the component diffusivity and migration coefficients by inserting the correlations from table \ref{tab:LiPF6EMCprops} into equations \eqref{LiPF6DandChirelations} and \eqref{LiPF6xirelation}, equations \eqref{binaryM_nu}-\eqref{binaryM_zz} were used locally to assemble the matrix $\textbf{M}_{Z}$. Simulations were performed assuming an equilibrium electrolyte composition of $\left\langle y_{\text{e}} \right\rangle = 0.15$ (ca. $1.9~\textrm{mol}\,\,\textrm{L}^{-1}$).

To model potentiostatic electrodes with fast heterogeneous reaction kinetics, the voltages at the positive electrode (p) and negative electrode (n) were respectively assumed to be $\text{V}^{\text{p}} = 10~\textrm{mV}$ and $\text{V}^{\text{n}}  = 0$. At each electrode surface, we applied the Robin boundary condition 
\begin{equation}
- \left. \left( \vec{i} \cdot \vec{n} \right) \right|_{k} = \frac{F z_+ i_{0}}{RT}  \left( \text{V}^{k}  - \left. \Phi_Z^0\right|_{k}  \right);
\end{equation}
here $\vec{n}$ is an outward unit normal vector, anodic overpotentials are positive, and $k \in \left\{ \textrm{p},\textrm{n} \right\}$. In simulations the exchange-current density $i_{0}$ was chosen to be very large ($1~\textrm{A}\,\textrm{cm}^{-2}$) so as to ensure that the solution-phase electrode potentials $\Phi_\textrm{Z}^0$ matched $V^{\textrm{a}}$ or $V^{\textrm{c}}$ to an appropriate degree of precision ($4$ digits). To relate the electrode currents to interfacial fluxes, the flux of lithium ions was taken to be
\begin{equation} \label{Boundarycondition}
\left. \left( \vec{N}_{\text{Li}^+} \cdot \vec{n} \right) \right|_{k} =  \frac{ 1 }{Fz_+} \left. \left( \vec{i} \cdot \vec{n} \right) \right|_{k} 
\end{equation}
at electrode surface $k \in \left\{ \textrm{p},\textrm{n} \right\}$; all other boundary conditions were set as homogenous Neumann conditions. 

The stripping and plating of lithium at the boundary as expressed in \eqref{Boundarycondition} implies a transfer of mass across the solution. It would therefore be inconsistent to take the reference velocity as the mass-average velocity and then assume this to be zero. Unfortunately, it is also not possible to solve for the distribution of mass-average velocity in two dimensions, because closure of problems in systems with more than one spatial dimension requires consideration of momentum continuity \cite{LIU2014447}. Instead we follow Newman and choose the reference velocity to be that of the solvent, and thence assume that velocity to equal zero uniformly. Practically, this amounts to a formal neglect of the Faradaic convection phenomenon. 

\begin{figure}
    \centering
    \includegraphics[width=8.5cm]{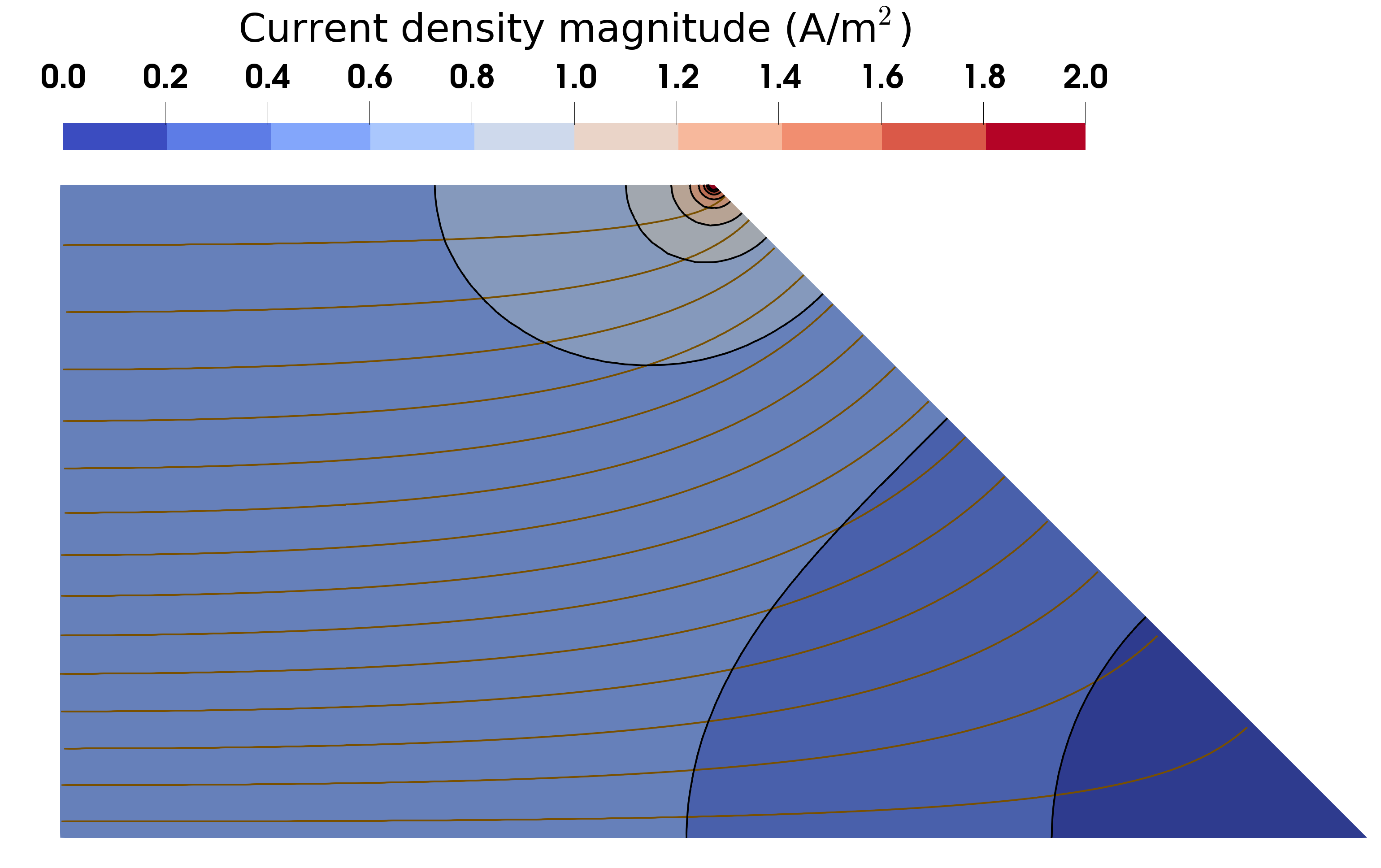}
    \caption{Steady-state current density $|\vec{i}|$ within a Hull cell with reversible electrode reactions subjected to a constant voltage drop. The streamlines denote trajectories of current flow from the positive electrode (left) to the negative electrode (right).}\label{fig:Hull cell}
\end{figure}

A steady--state simulation of the potentiostatic Hull cell was performed with Firedrake software \cite{Rathgeber2016}, using the MUMPS direct linear solver \cite{MUMPS01, MUMPS02} via PETSc \cite{petsc-user-ref, petsc-efficient}. The mesh of the geometry was constructed using the Gmsh software \cite{GMSH}. Each linear system had $183,831$ degrees of freedom. The finite elements were chosen as $\text{CG}_{2}$ for component mole fractions/voltage and $\text{DG}_{1}$ for fluxes/current. The computed current distribution is plotted in Figure \ref{fig:Hull cell}. Observe that the current concentrates at the upper corner because of the obtuse angle with an insulating surface, and is dispersed at the lower corner, where the angle is acute, as might be expected from electrostatic analysis of the wedge geometry in cylindrical coordinates \cite{newman2004electrochemical}. 

This example illustrates the potential of the structurally electroneutral Onsager--Stefan--Maxwell framework to model concentrated electrolytes with full composition dependence of all independent thermodynamic and transport properties, on complex geometries. Similarly we might consider modelling molten salts, or cosolvent electrolytes, or transient dynamics of any of these. Subject to availability of the necessary property data, the scope of applications can be extended much further. 

\section*{Conclusion}

We have detailed how local electroneutrality affects Onsager--Stefan--Maxwell mass transport within electrolytic solutions. Choosing a salt--charge basis according to the process we outline in section \ref{sec:construction} allows one to establish constitutive equations for component chemical-potential gradients in terms of a symmetric thermodynamic-factor matrix. Crucially, the use of a salt--charge basis also obviates the need to impose local electroneutrality as an algebraic constraint on the electrolytic mass-transport problem, allowing it to be woven into the parameter space of the differential governing system. This leads to the electroneutral Onsager--Stefan--Maxwell equations \eqref{OSM_neutral}, which phrase $n-1$ component chemical potential gradients and the salt--charge potential gradient explicitly in terms of $n-1$ uncharged component fluxes and current density. Salt--charge potential relates in a straightforward way to the electric potential measured by a reference electrode of a given kind. 

A process was established to invert the electroneutral Onsager--Stefan--Maxwell laws into the flux-explicit form familiar from concentrated solution theory, allowing the identification of standard transport properties such as conductivity, component diffusivity, and species transference numbers. We introduced migration coefficients, a useful alternative that remedies the issue that transference numbers as commonly understood cannot describe all the electro-osmotic interactions in systems containing multiple uncharged species. Several examples were given to substantiate how the salt--charge basis concept enables the direct formulation of flux-explicit constitutive laws for mass and charge transport in locally electroneutral concentrated electrolytic solutions. 

Establishing a salt--charge basis allows one to formulate the transport equations for locally electroneutral electrolytic solutions in a way that retains the spectral structure afforded by the Onsager--Stefan--Maxwell laws for nonelectrolytes. The utility of this perspective was underlined by using a code designed for nonelectrolytes to simulate electroneutral transport within a nonideal binary electrolytic solution with composition-dependent properties in a Hull cell.

\section*{Acknowledgements} 

This work was supported by the Engineering and Physical Sciences Research Council Centre for Doctoral Training in Partial Differential Equations: Analysis and Applications (EP/L015811/1), the Engineering and Physical Sciences Research Council (grants EP/R029423/1 and EP/W026163/1); the Clarendon fund scholarship; and the Faraday Institution Multiscale Modelling project (subaward FIRG025 under grant EP/P003532/1).


\bibliography{bib.bib}

\end{document}